\documentclass[fleqn,12pt]{wlscirep}
\usepackage[utf8]{inputenc}
\usepackage[T1]{fontenc}

\usepackage{multirow}

\newcommand{ \pa }{\textcolor{black}}

\title{Prediction and experimental evidence of the optimisation of the angular branching process in the thallus growth of {\it Podospora anserina} }

\author[1]{Clara Ledoux}
\author[1]{Florence Chapeland-Leclerc}
\author[1]{Gwenaël Ruprich-Robert}
\author[1]{Cécilia Bobée}
\author[1]{Christophe Lalanne}
\author[1,*]{Éric Herbert}
\author[1]{Pascal David}
\affil[1]{\pa{Université Paris Cité, CNRS, UMR 8236 – LIED, F-75013 Paris, France. }}

\affil[*]{Corresponding author, \pa{eric.herbert@u-paris.fr}}

\begin{abstract}
{Based upon apical growth and hyphal branching, the two main processes that drive the growth pattern of a fungal network, we propose here a two-dimensions simulation based on a binary-tree modelling allowing us to extract the main characteristics of a generic thallus growth.  In particular, we showed that, in a homogeneous environment, the fungal growth can be optimized for exploration and exploitation of its surroundings with a specific angular distribution of apical branching. Two complementary methods of extracting angle values have been used to confront the result of the simulation with experimental data obtained from the thallus growth of the saprophytic filamentous fungus {\it Podospora anserina}.}
Finally, we propose here a validated model that, while being computationally low-cost, is powerful enough to test quickly multiple conditions and constraints. It will allow in future works to deepen the characterization of the growth dynamic of fungal network, in addition to laboratory experiments, that could be sometimes expensive, tedious or of limited scope.
\end{abstract}

\begin{document}
\maketitle

\section*{Introduction}

\noindent
{Filamentous fungi are characterized by their ability to form an interconnected hyphal network, the mycelium, based upon some key cellular processes, {\it i.e.} hyphal tip (or apex) growth, branching and hyphal fusion (also known as anastomosis), thus conferring a great flexible morphology and a remarkable capacity of adaptation to very diverse ecosystems.\cite{fricker_mycelium_2017} In particular, saprophytic fungi, known as short-range foragers, evolves in a highly competitive habitat and thus appears to be highly challenged by resource-limited and patchy environment, combined to a fierce competition with other organisms.\cite{sanati_nezhad_quantification_2013} Then, these fungi must adapt their growth and have always to find a compromise between the need to occupy the space potentially threatened by other organisms (colonization) and the need to optimally draw resources from where the thallus has developed (densification). Previous studies have reported that hyphal tips at the biomass edge could be associated with exploration while hyphal tips behind the biomass front could be most associated with resource exploitation.\cite{boswell_functional_2002} In the same way, \cite{lew_how_2011} described the apical cells (or leading hyphae) from the fungal network as the first cells to invade new territory and are generally engaged in nutrient acquisition and sensing of the local environment. However, until now, the way a fungus optimizes its growth through an efficient compromise between the maximization of the surface occupancy and the increasing production of length is still unclear. 
\\
For many years, \pa{a review can be found in,\cite{Heaton2012Apr}} mathematical modelling and measurement of specific quantitative parameters through the use of image analysis have greatly contributed to a better understanding of fungal network expansion and topology.  Usually, these approaches attempt to explore the fungal development in two or three dimensions, in order to define some specific macroscopic and microscopic observable and to describe, explain and predict development of the fungal network in various more or less constraining environments. 
It was shown in\cite{boswell_functional_2002} that, in the saprophytic fungus {\it Rhizoctonia solani}, translocation (nutrient transport) was mainly diffusive in homogeneous environment, when growth experiments were compared to a detailed mathematical modell; in the latter case, translocation was considered to have both diffusive and metabolically-driven components. Falconer {\it et al.} \cite{falconer_biomass_2005} demonstrated that fungal phenotype could be modelled as an emergent phenomenon resulting from the interplay between local processes involved in nutrient uptake and remobilization of internal resources, and macroscopic processes associated with their transport. 
In order to enlarge the spatio-temporal scale of the hyphal growth of the brown rot fungi {\it Postia placenta}, \cite{du_lattice-based_2018, du_3-variable_2019} proposed a lattice-based system derived from the biological mechanisms of hyphal development and then a 3-variable Partial Differential Equation (pde) model for predicting fungal growth derived from microscopic mechanisms. 
As highlighted in \cite{boswell_linking_2008}, lattice-based modelling allows for the inclusion of fundamental biological processes of the thallus growth, as anastomosis, branching or translocation while maintaining a decent computation duration. However, even if the regular imposed geometry can be of different form but is expected to impact the global thallus geometry and consequently its  functionality, \textit{i.e.} the comportment and interaction (effects) with the environment. It is then desirable to develop a lattice-free approach, in which the thallus is described as a collection of line segments. The lack of constraint on the hyphal growth orientation dramatically increases the computational complexity and can be cumbersome to implement. These lead to ignore crucial processes like anastomosis (see for example \cite{meskauskas_simulating_2004}). 
More recently, using high computational resources, \cite{vidal-diez_de_ulzurrun_modelling_2017} described a three-dimensional lattice-free fungal growth model that is able to simulate both the biological processes driving fungal growth and the hyphal response to environmental stimuli, and that can be compared to growth features obtained {\it in vitro}.

\noindent
Recently, we have developed an automated and reproducible experimental device to track the hyphal network construction of the filamentous saprophytic fungus {\it Podospora anserina} constrained to grow on the planar surface of a Petri dish, in a standard environment of {\it in vitro} growth.\cite{dikec_hyphal_2020} Such a system allowed us to monitor time series of images of the fungal thallus. The acquisition step was followed by a robust image analysis process, leading to the extraction of a set of reproducible quantitative parameters, as the total length of the mycelium, the number of nodes or vertices ({\it i.e.} hyphal fusion and branching points), and the number of apexes. Such a systematic spatial and temporal exploration enabled us to estimate a set of key physiological features of the fungal network, such as the branching dynamics and more generally the spatio-temporal patterns of the network. 
\\
\noindent
This biological system is obviously far from the \textit{in vivo} growth of a saprophytic fungus such as {\it P. anserina}. However, it constitutes an excellent starting point to describe the thallus growth using mathematical concepts and languages. Indeed, the objective of our mathematical modelling is to reduce a complex biological system to a simpler model, which is able to partly reproduce, or even better predict, the real system. A recurrent question is then to find the optimal degree of simplification for such a model, which should be neither too simple to avoid straying from realistic predictions, nor too complex to solve using numerical methods.\cite{du_lattice-based_2018}
We report in this work a two-dimension lattice-free model that aims to capture the main characteristics of the thallus growth. This model relies on a binary tree growth with short computation time, and it allows for testing multiple hypotheses and configurations quickly.
As it is commonly accepted that the growth pattern of mycelium results from two main processes, apical growth and branching,\cite{fricker_mycelium_2017} we focused on these two parameters to simulate the fungal growth.  
We propose to deepen the characterization of the growth dynamic of the {\it P. anserina} fungal network through the construction of a growth model for this fungus and then to confront these simulation approaches with angle measurements. Two complementary methods for extracting angle values have been used. The first one consists in direct empirical measurements of angles from apical branching based on a selected collection of nodes. The second one relies on a Geographic Information System (GIS) approach in which all nodes are considered (apical and lateral branching). 
Our results allow us to compare the distribution of apical angular values with the two competing but vital processes of densification and extension, on a standard culture medium, assumed to be homogeneous and optimal for the fungal growth. We can thus show that far from being random, they correspond on average to an optimal configuration. Beyond this work, the final model presented here could be a powerful tool in future studies for predicting the mycelial growth over a larger spatial scale and under various and more fine-grained conditions or constraints than most {\it in vitro} experiments could afford.}

\begin{figure}[htbp]
\centering
\includegraphics[width=0.5\textwidth]{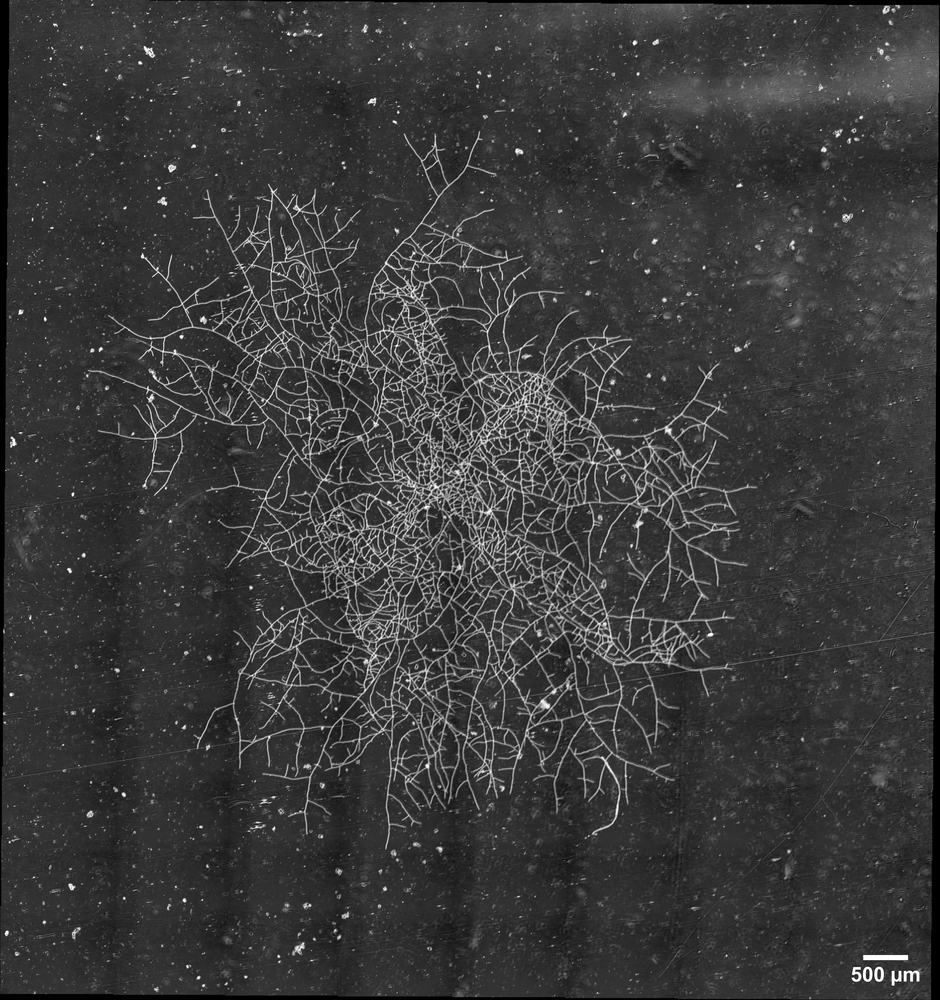}
\caption{Thallus of \textit{P. anserina} reconstructed from $8 \times 14$ tiles, extracted from experiment \cite{dikec_hyphal_2020} 15 hours past the ascospore germination (see the text and Tab.~\ref{tab:exponent} for more details).}
\label{fig:p50}
\end{figure}


\section*{Simulation}
\subsection*{Network topology}

\noindent
There are different types of complex networks allowing for the characterization of branching network:\cite{newman_structure_2006, v_latora_complex_nodate} $i)$ {\it Random networks} for which the connections between the nodes are random and are free from the physical distance that separates them $ii)$ {\it Totally interconnected networks} which allow all the nodes to be linked together $iii)$ {\it Freescale networks} which are characterized by a number of connections between nodes which follows a power law ({\it i.e.} many nodes which have few connections but some nodes which have many) $iv)$ {\it Smallworld} type networks which are characterized by strongly interconnected subnets, which are themselves connected to each other via a few privileged links. It is also necessary to distinguish between highly centralized networks, which assume privileged nodes in order to transmit information from one node to another (the geodesic in terms of information transmission is then not optimal), and networks without privileged centers that allow an optimal transmission of information. A third important aspect is the existence of networks that do not have spatial dimensions (internet for example) and others whose spatial dimension is critical in the sense that the probability of connection between two spatially distant nodes is zero. Finally, we must distinguish between static networks (in the sense of the number of nodes and not of the activity that takes place there) and dynamic networks which are capable of complicating and increasing their number of nodes and links as a function of time.

\noindent
The study of {\it P. anserina} in~\cite{dikec_hyphal_2020} allowed to identify the core characteristics the simulation should be based on. {\it P. anserina} thallus is a spatialized and dynamic network (Fig.~\ref{fig:p50}). Accordingly, the probability of connecting spatially distant nodes (or vertex) is zero. However, if the thallus of {\it P. anserina} seems to be a centralized network in its first phase of growth, the branching process dramatically and quickly increases the network complexity, meaning that the number of possible paths between the different regions of the thallus become very high. The centrality of the network is then lost, since the information can be exchanged without going through a center. Moreover, a partial destruction of the initial center (the ascospore location) does not affect the ongoing development of the network.\cite{fricker_mycelium_2017}  The observation of {\it P. anserina} network shows nodes (or vertices) which are generally connected to three other nodes that can be distinguished by their chronological growth~: there are indeed one earliest node, or {\it mother node}, and two {\it daughter nodes}. An illustration can be found in Fig.~\ref{fig:schema}.  \\
{\it P. anserina} network can be distinguished from the four major types of networks previously mentioned. If this network was not a dynamic network as it was defined above, at a given point in time, {\it P. anserina} network could be locally related to a {\it hierarchical and modular tree} where modularity (existence of a basic architecture) is important, the heterogeneity (variance of the distribution of the number of links) is around zero and whose randomness (probability of creating links between the nodes) would be very low.\cite{sole_information_2004} It is therefore possible to model the growth of {\it P. anserina} as different networks linked together by causal relationships and spatial constraints. 

\noindent
To implement this simulation we made the following assumptions: $i)$-the environment is supposed to be homogeneous (in terms of resources, light, and other exogenous constraints ...) and invariant over time; $ii)$-the growth conditions are standard (temperature, pressure, etc.) and invariant over time; $iii)$-there is no limits of deployment in space and time; $iv)$-the initial state is defined as the germination of a single ascospore.
These assumptions are not restrictive. Except for hypothesis $iii)$ they correspond to the experimental conditions. 
\\
\noindent
Basically, a simulation can be viewed as a two-step process: \\
-- First, the {\it generation step} corresponds to the theoretical growth of the network ({\it i.e.} the theoretical law). This step is built from assumptions on probability laws and their parameters. \\
-- Second, the {\it detection step} corresponds to the observation of this growth. This step is also built on probability laws but only relating to the observation process. It must reproduce as well as possible the entire experimental process. \\
This two-step process is justified by the fact that we never observe or measure what is ``true'' but only its convolution by the whole observation chain which can possibly generate biases, and necessarily has some resolution.

\subsection*{Generation}
The aforementioned considerations on the kind of network generated by {\it P. anserina} lead to choose a growth model in the form of a {\it full binary tree} (or {\it proper binary tree}).\cite{garnier_discrete_2009} In this model, each branch (hypha) is driven by its tip (apex) that in turn is allowed to divide into a pair of sub-branches. This process is called \textit{apical branching} in this paper, in line with the dynamic growth vocabulary. When apical branching occurs, the distribution of branch lengths and that of the angles between newly created branches with respect to the mother branch follow differentiated probability laws. 
An apical branching process is characterized by the emergence of a 3-body vertex, denoted $V_3$, which is the point of intersection of the 3 branches and two 1-body vertex, denoted $V_1$, the apexes of the branches (Fig.~\ref{fig:schema}). Of course, for the duration of binary tree growth, both number of $V_3$, $N_{V_3}$, and of $V_1$, $N_{V_1}$, depend on the instant of observation and are related by $N_{V_3} = N_ {V_1}-(N_{ger}-1)$ where $N_{ger}$ is the initial number of branches emerging from the ascospore, or $N_{V_3}(t)\approx N_{V_1}(t)$ if $N_{V_3}(t)$ is large enough (or equivalently that the duration of growth is large enough). In the following, we will assume this condition is always verified. \\
\noindent
\pa{The length of the branches is a random variable} assumed to be correctly described using a Gamma law which is a continuous and positively skewed distribution. Moreover,  other laws of probability can be obtained by changing parameters of the Gamma law:

\begin{align}
    G(x|k,\theta)&= \frac{x^{k-1} \, e^{\frac{x}{\theta}}}{\Gamma(k) \, \theta^k}
\end{align}
\noindent
with $x$ the length, $k$ the scale factor, $\theta$ the shape factor, and $\Gamma(k)$ the Euler function. 
Using a particular choice of parameter values for $k$ and $\theta$, the Gamma law allows to fall back to $i)$-the exponential law ($k=1$, $\theta=1/\lambda$) $ii)$-the Maxwell-Boltzmann's law ($k=3/2$, $\theta=2a^2$) $iii)$-the $\chi^2$ law ($k=\nu/2$, $\theta=2$ with $\nu$ the number of degrees of freedom, d.o.f). Consequently, the Gamma law offers a lot of freedom to generate the length of the branches. \\
We considered a Gaussian probability law for the angles: $ N(x|x_0,\sigma)$ with $x$ the angle, $x_0$ the mean and $\sigma^2$ the variance. Because it was empirically observed that apical branching leads to two different angles, the parameter relating to the mean, $x_0$, is different for the two branches coming from a $V_3$. The variances for the angles are chosen according to experimental data. The position of the branches in relation to the direction of the mother branch - to the right or to the left - is managed by a probability ({\it i.e.} Bernoulli law). So, it is impossible to force a particular direction for one or both forms in order to break the chiral invariance of the growth. 

\noindent
We call {\it operating branch} the branch coming from a vertex $V_3$ which has the most important angle (the angle is noted $\theta_o$) with respect to the segment that connects the vertex to the vertex from which it came (mother vertex) and we call {{\it exploratory branch} the branch with the smallest angle (the angle is noted $\theta_e$) (see Fig.~\ref{fig:schema}).
This nomenclature is justified by the fact that the two branches do not have the same scope in terms of network growth: the branch showing the smallest angle tends to perpetuate the direction of the mother branch in order to explore the environment and then to capture new resources, while the other branch occupies space inside the network in order to optimally draw resources from the neighborhood.
\begin{figure}[htbp]
\centering
\includegraphics[width=0.65\textwidth]{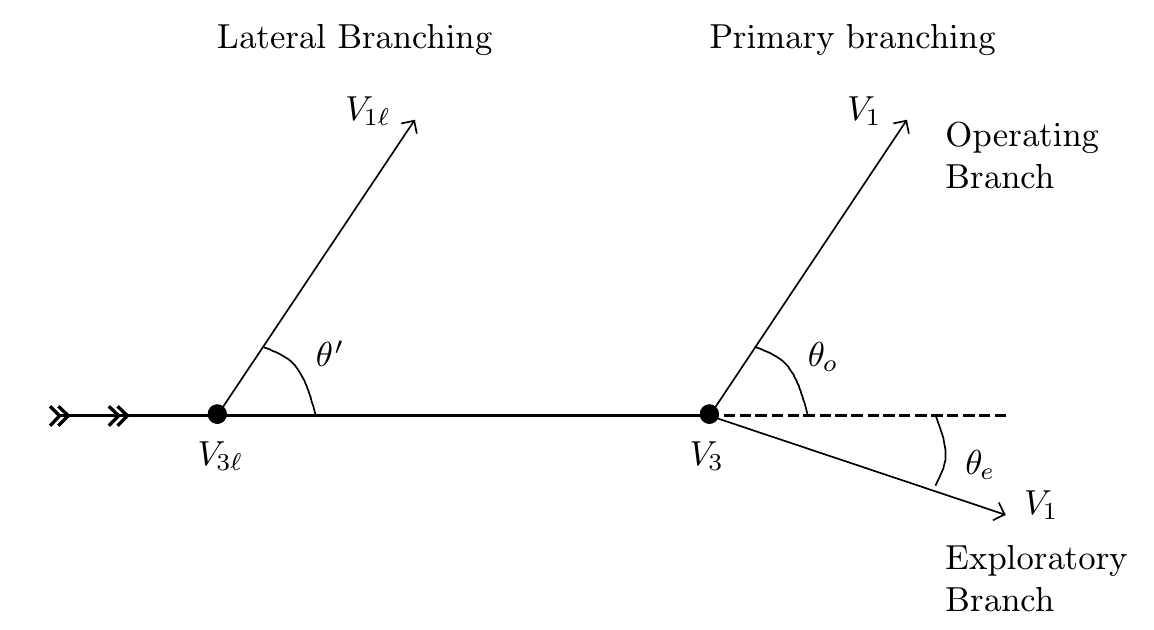}
\caption{Definitions used in this work. The thick line is the mother hypha, while the thin lines are the daughter hyphae. We consider two types of 3-degree vertices. $V_{3\ell}$ are lateral branchings, connected with the angle $\theta'$. $V_3$ are apical branchings on which two branches are connected: an operating and an exploratory branch with respective angles $\theta_o$ and $\theta_e$.  $V_1$ (resp. $V_{1\ell}$) are 1-degree vertices (\textit{i.e.} apexes) coming from $V_3$ (resp. $V_{3\ell}$).}
\label{fig:schema}
\end{figure}

\noindent
Experimental observations show another type of branching process giving rise to another type of 3-body vertices: interior vertex leading to the creation of lateral branch}. A branching occurring on a segment bounded by two pre-existing $V_3$ vertices, or a $V_3$ and a $V_1$ vertices can give a new lateral branch at a later moment during the growth process. To distinguish between these branches we call them {\it lateral (operating) branch} and the 3-body and 1-body vertices attached to them are denoted respectively $V_{3\ell}$ and $V_{1\ell}$ see Fig.~\ref{fig:schema}). \\
The creation of a {\it lateral branch} is also driven by a probability law. This law is unknown (as well as its parameters) leading to an important difficulty in implementing the phenomenon. However, we choose a power law based on two considerations : $i)$-the duration that separates the observation from the creation of the two vertices bounding the possible lateral branch $ii)$-the length between the pre-existing vertices. In this case, the probability of generating a lateral branch as a function of the length and time parameters could be written: 

\begin{align}
p &= p_0 \left(\frac{x}{x_0} \right)^\alpha
\label{eq:proba_lateral}
\end{align}

\noindent
where $x$ is the length of the branch ({\it i.e.} the length between the two vertices), $x_0$ a length parameter ($x_0 \geq x$) ({\it i.e.} the greatest length generated by the probability law, typically the mean ($k \theta$) plus $3$ times the standard deviation ($\sqrt{k} \theta$)), $\alpha$ a parameter of the power law, typically $\alpha \sim 2$ and $0 \leq p_0 \leq 1$ a scale parameter to fix the probability, typically $p_0 \sim 1/2$.
The duration between the observation and the creation of the two vertices is not apparent in equation~\ref{eq:proba_lateral} because this probability is recomputed after each generation of newly generated branches (of hypothetical duration). \\
The location of the lateral vertices $V_{3\ell}$ between two vertices follows a uniform law. However we implement a censorship zone near the vertices so that the uniform distribution does not have the vertex-vertex distance as a support. Empirically, we see that the angular distribution between the segment connecting the two vertices and the {\it lateral operating branch} is about the same as the angular distribution of the {\it operating branch}. Therefore we choose the same law with the same parameters. Likewise, the length of the branches thus created follows the same probability law as that of the {\it exploratory branches} and {\it operating branches} (with different parameters). The position of the {\it lateral branch} in relation to the direction of the mother branch - to the right or to the left - seems to be in the direction of the curvature of the hyphae. Also, the chirality for these branches can be broken by a probability parameter.
The control of these laws and their parameters is a challenge which is raised by the study of the calibration of simulation to the data (see the next section). 

\noindent
The curvature between any two vertices 
is not taken into account in the present work for the following reasons: $i)$-The curvature can not be related to anything particular during the growth process (to a first approximation); $ii)$-The chiral symmetry can be broken from the angular distributions if one wishes to obtain a global curvature effect. In this sense, the chirality breaking of the {\it lateral branches} takes into account the curvature of the segments between the points formed by the triplet, grandmother, mother and daughter -- two adjacent segments; $iii)$-From the experimental data it can be seen (not shown) that mean curvature (\textit{i.e.} the curvature over the different fractions of the length between two vertices) is locally zero. 

\noindent
Because the experimental process introduces a time dependency through the reconstruction of network images and because we want to study the growth dynamics of the network, we define a time between vertex generations. At this step the time is given by number in generation agency between daughter-vertices and mother-vertex. For the {\it lateral branches} the time is given by the relative moment - in terms of generation - at the creation of the branch. However, this theoretical time will be scaled by the data (see calibration section) and we will take into account a possible growth interruption during apical branching process as well as differentiated apex velocities of {\it exploratory}, {\it operating} and {\it lateral} branches. 
For the simulation step, however, these velocities are given by a Gaussian probability law with average values and a small shape (standard deviation) and we assume these velocities are constant during growth. 

\noindent
To sum up, the generation time found is certainly not the growth time of the fungus. However, this time is necessarily proportional to the growth ``true'' time. The constant of proportionality must be obtained from calibration on the data. 

\begin{figure}[htbp]
\centering
\includegraphics[width=0.65\textwidth]{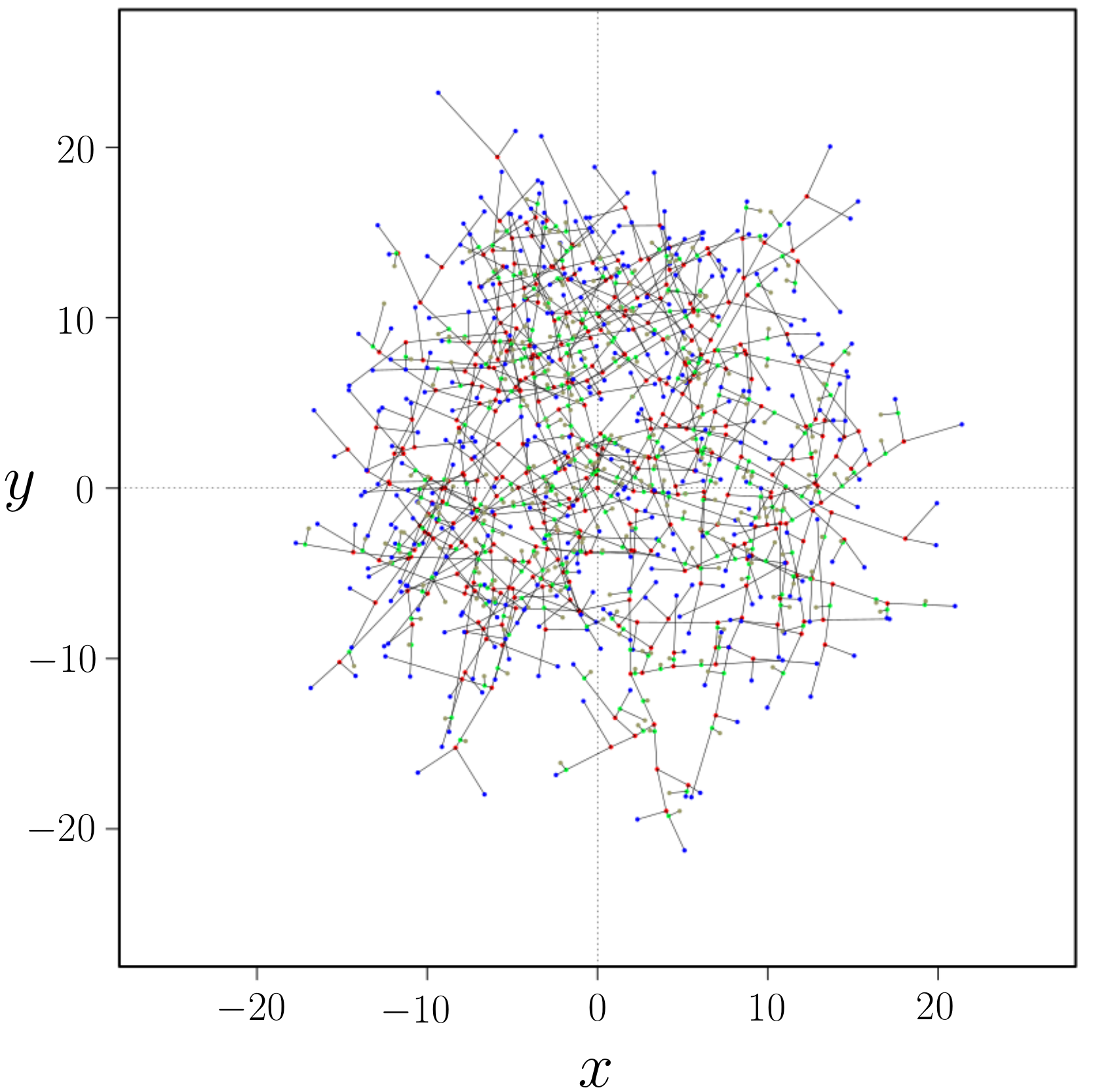}
\caption{Final state (after nine generations) of simulated network for a set of standard parameters. Apical branching led to  $N_{V_1}=384$ apexes (shown in blue) and $N_{V_3}=383$ vertices (in red). Lateral branching led to $N_{V_3\ell}=295$ vertices (in green) and $N_{V_1\ell}=295$ apexes (in grey). The spatial dimension in the representation is arbitrary.}
\label{fig1}
\end{figure}

\noindent
Simulation can start with as many branches as desired. We make this number coincides with the empirical observation of the ascospore germination that shows three initial branches with an angular separation of approximately $ 2\pi/3$. We then apply the growth as a binary tree for each of three branches as described below.
Fig.~\ref{fig1} gives an example of the final state of growth. 
\subsection*{Detection}
The result of the measurements (or in an equivalent way, of the observations) always corresponds to a convolution of the reality, the \pa{probability density function (pdf)} of ``true value'', with the acquisition chain (the pdf of resolutions). In order to make numerical results comparable to the experimental observations, it is then necessary to introduce the same bias and measurement uncertainties. In what follows, \textit{observations} may refer to experimental or numerical data.
There are different interferences which confuse the values of the quantities of interest that may depend on the entire detection chain which goes from the shooting to the post processing of the digitized image.
\begin{itemize}
\item First, the experimental observation process is not able to distinguish ``true'' $V_3$ vertices ({\it i.e.} from fission process or equivalently branching, $V_3$ and $V_{3l}$) from geometric vertices ({\it i.e.} geometric intersection of two branches because of two-dimensional observation). The number of these geometric vertices depends on the time of observation (because of the creation of {\it lateral operating} branches and {\it operating} branches). Indeed, the probability of obtaining branch intersections increases with the density of branches which also increases with time. \\
We distinguish two situations: $i)$-either the intersection is purely geometric (the experimentally observed vertex may result from the superposition of two branches. 
These vertices are noted $V_{3i}$ because even this kind of vertex shows four branches. Usually, network reconstruction based on experimental images in this situation  leads to two vertices with 3 branches in close proximity among the four branches constituting the intersection. In the second situation $ii)$ it gives rise to a merger. We know that a process of hyphal fusion ({\it i.e.} anastomosis) can occur in {\it P. anserina}. These vertices are noted $V_{3o}$ despite the fact it still is a vertex with four branches. The anastomosis phenomenon arises following an unknown probability. However, the number of vertices resulting from the hyphal fusion process is certainly very low compared to the number of $V_3+V_{3l}$ and $V_{3i}$ vertices. Even if it is marginal, this hyphal fusion phenomenon allows a more efficient information transition within the network and it also makes it possible to be freed from a dependency in the centrality of the network. It is possible to qualitatively study, via simulation, the gain of a hyphal fusion phenomenon by building the {\it adjacency matrix} \cite{newman_structure_2006, v_latora_complex_nodate} and by iterating on it so as to connect all the vertices to each other in a minimum time (or distance).  \\
The presence of these vertices blurs the information on the ``true'' number of vertices $V_3$, however, since $N_{V_3} \approx N_{V_1}$ and $N_{V_{3l}} = N_{V_{1l}}$ for a long duration, we get a very reasonable order of magnitude of $N_{V_3}+N_{V_{3l}}$. \\
This limitation of experimental observation forces us to abandon the idea of working on historical markers of the development of a thallus. Indeed, if the vertices $ V_1 $ give us an image of the current state of development, the vertices $ V_3 $ - resulting from the $ V_1 $ which are only intermediate stages - constitute the deep history of growth. In order to remedy to this corner case, we developed a method to clearly identify geometric intersections. 
\item A second effect is the resolution. On the one hand, the optic we use obviously has a finite resolution. On the other hand, there are experimentally limits in the reconstruction of the images. We take these resolutions into account by convoluting the position of the different vertices positions and branches with a Gaussian resolution. Although taken into account in the simulation of the detection, this phenomenon is not of capital importance in the observation of the macroscopic characteristic quantities of the growth of the network. 
\item The third effect is also due to the finite resolution of the optic. A vertex $V_1$ or $V_{1\ell}$ can be very close to a branch. In this case it is impossible to distinguish it from the branch. This situation generates a vertex that behave as in the case of geometric intersections. We do not need to distinguish them from the vertex coming from intersections ({\it i.e.} $V_{3i}$), and so we note them $V_{3i}$. It is easy to see thanks to the simulation that this phenomenon only becomes important when the density of hyphae is important, therefore for very long growth times.
\item Finally we also take into account the possibility of having {\it ghost} vertices on the images coming from the optics or from the reconstruction of the images. We note this vertices $V_{3g}$ or $V_{1g}$. This type of vertex is minimicked by specific processing during image reconstruction. Also, its number should be marginal and does not depend on time.  Here again, we can experimentally verify that this process is marginal. 
\end{itemize}
The number of numerically observed vertices, $N_{V_{3ob}}$ and $N_{V_{1ob}}$, is therefore the sum of the different contributions of the effects listed above, while dominant parasitic vertices are mostly of geometric nature. 
It is difficult to build a theoretical model which allows to obtain this number even if we know that this number grows very significantly with the time of observation (typically exponential growth). The simulation makes it possible to obtain this number.
%

\subsection*{Calibration}
In the simulation there are several parameters that need to be calibrated on the data. In order to do that we present here two sensitive variables. 
\subsubsection*{Time scale}
To scale the time we use the number of vertices $V_{1ob}$ and $V_{3ob}$. 
It is much more complicated to find the law that governs the number of $V_{3ob}$ vertices than that of $V_{1ob}$ because of the presence of $V_{3i}$ whose law is unknown. On the other hand, because there are more intersections between branches than there are apexes near a branch, one would expect to obtain a larger contribution of branch-branch mergers, even if the merging process remains marginal (the $V_1$ contribution in $N_{V_{3o}}$). So, the calibration must be carried out on the vertex $V_{1ob}(t)$ and we will check that the number of vertices $V_{3ob}(t)$ follows the empirical law, see Eq.\,\ref{eq:fitALN}, from the data. \\
The number of $V_{1ob}(t)$ is formally : 

\begin{align}
    N_{V_{1ob}}(t) &= N_{V_{1}}(t)  + N_{V_{1\ell}}(t) + N_{V_{1g}}(t) +\epsilon_1  \,  N_{V_{3o}}(t) +\epsilon_1^{\prime} \,  N_{V_{3i}}(t)
\end{align}

\noindent
For reasonably long time and with a quality control procedure, we can neglect some terms of the sum, \textit{i.e.} $N_{V_{1g}}(t)$ is very small compared to other numbers of vertices.
The number of hyphal fusion apexes, $\epsilon_1 \,  N_{V_{3o}}(t)$, is a second-order corrective term because it depends on anastomosis which is a marginal process. The number of apexes-branches intersections  $\epsilon_1^{\prime} \,  N_{V_{3i}}(t)$ 
depends on the resolution of the optics and the reconstruction which can be controlled (it is always possible to degrade the resolution in order to check this number at a given time). However,  we cannot effectively neglect this component even if the vertex density is not very high (we will see on the data observed experimentally that it occurs beyond a certain growth time). Also the calibration on the experimental data we use can only be carried out in a growth time domain. \\
The term $N_{V_{1\ell}}(t)$ depends on the number of branches at the time $t-\delta t$ (with $\delta t$ the time between two observations in experimental data and index of generation in simulation) and the full probability of producing a lateral branch before this time. The number of branches at the time $t-\delta t$ is $ N_{V_{1}}(t-\delta t ) +N_{V_{3}}(t-\delta t )-2 \approx 2N_{V_{1}}(t-\delta t )$. Finally because we are dealing with the growth of a binary tree, $N_{V_{1}}(t)$ is know to be $N_{V_{1}}(t) = 2^{at}$ with $a>0$, where $N_{V_1}$ is a  scale parameter for the time. Using these approximations the number of $V_{1ob}(t)$ is then:

\begin{align}
    N_{V_{1ob}}(t) & \sim 2^{a \, t} + 2^{a \, (t-\delta t)} \, p \approx A \, 2^{a \, t}
\end{align}
\noindent 
with $\delta t/t \leq 1$ and with $p$ the mean fraction of lateral branches for a branch (which is a function of the probability to create  \textit{lateral branches}) which is constant by hypothesis. $A$ is a constant amplitude which takes into account $p$ and the fact that if $\delta t$ is perfectly defined, the origin of the times is not: there is a ``latency'' time between the start of germination and the first observation. Indeed, we know that for $t = 0$ we must obtain $ 0 \leq N_{V_{1ob}}(0) \leq 3$. We can then rewrite this number of vertices as $N_{V_{1ob}}(t) = A^{\prime} 2^{a(t+t_0)}$ and in this case $0 \leq A^{\prime} \leq 3$ and $t_0$ is the ``latency'' time. So, $A^{\prime} 2^{at_0}=A$. \\

\begin{figure}[htbp]
\centering
\includegraphics[width=0.45\textwidth]{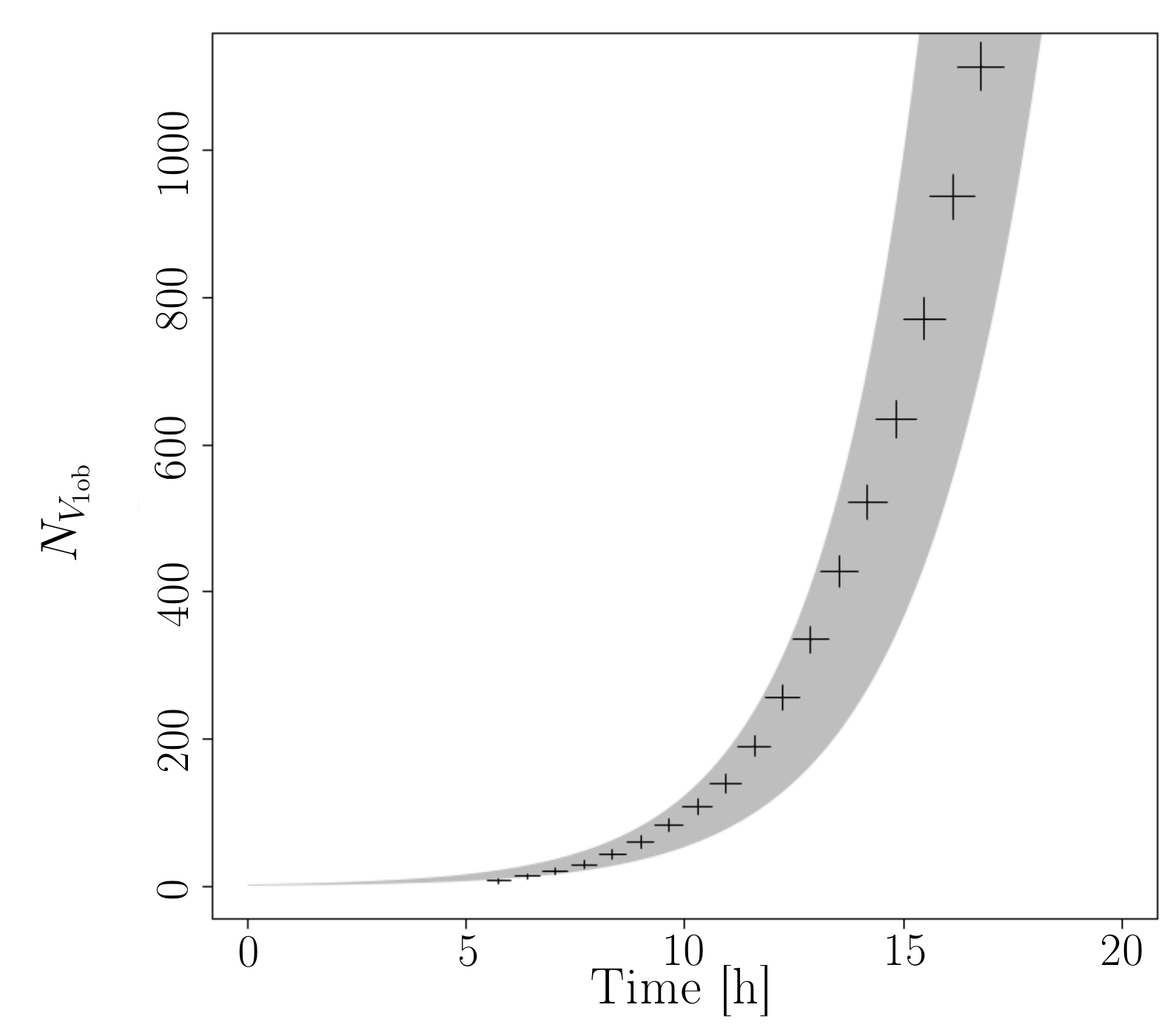}
\includegraphics[width=0.45\textwidth]{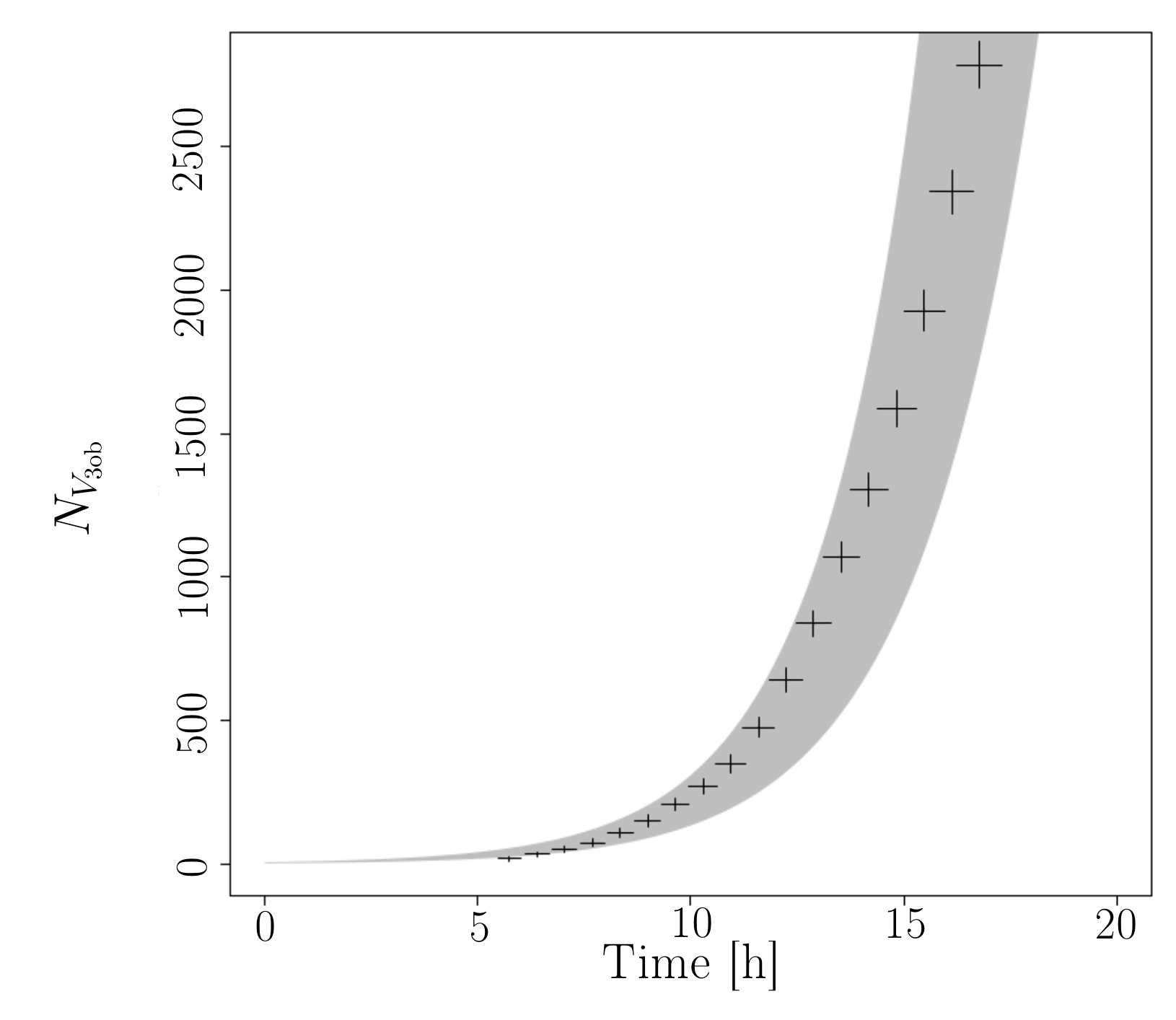}
\caption{Left, one-standard deviation range of the experimental data fit of $V_{1ob}$ (grey) versus time, points: number of $V_{1ob}$ coming from the simulation (black) with scaled time. Right, one-standard deviation range of the experimental data fit of $V_{3ob}$ (grey) versus time, points: number of $V_{3ob}$ coming from simulation (black) with scaled time to the data obtained from $V_{1ob}$. See text for details.}
\label{fig4c}
\end{figure}

\noindent
We fit experimental data with this law in order to get the 2 parameters (see experimental approach) then 
we fit $N_{V_{1ob}}(t)$ coming from the simulation with the same law. In order to calibrate the simulation time we get the relation $(A 2^{at})_{data}=(A 2^{at^{\prime}})_{sim}$ and compute $t^{\prime}$ with the hypothesis that $t^{\prime}$ is a linear function of $t$. \\
In the \pa{Fig.~\ref{fig4c} (left) }, we plot the uncertainties for one standard deviation obtained from the fit parameters of data (grey area). We scale the time (and compute one standard deviation for this time from the uncertainties on the parameters of the 2 fits carried out) and plot the points coming from the simulation (the black cross, the uncertainties on the numbers of $V_{1}$ coming from simulation follow a Poisson distribution). \\
The law for the number of $V_{3ob}$ is $ N_{V_{3ob}}(t) \approx 2^{at}\sum_n c_n t^n$ because $N_{V_{3}}$ have the same law as $N_{V_{1}}$ since  
$N_{V_{3i}}$ is the dominant term of $N_{V_{3ob}}$ and we assume a regular increase with time for this number. In practice, the series can be limited to the first 2 terms (check the first negligible order in the fit). Grey area in the Fig.~\ref{fig4c} (right) show one standard deviation giving from the fit parameters of data.
If the scale factor obtained from $V_{1ob}$ is correct, the points from the simulation for the 3-bodies vertices,  must be inside a side band from the data.  \\
The result shows a good agreement. We conclude that i) we have correctly defined the time scale factor, and ii) binary tree growth is the right model for dynamic growth of {\it P. anserina}. However, it is certain that the law of growth of the fungus we use, $A\,2^{a \, t}$ ($a>0$), diverges when time tends towards infinity. This behaviour is certainly not permissible. Also, the law actually followed by this growth should rather be of the form $N_{V_{1ob}}(t) = \frac{1}{b+c \, 2^{-a \, t}}$
with $a,b,c$ real positive parameters. Using this law, we can see that if $at$ is not too big and if $c\gg b$, the law becomes  $N_{V_{3o}} \approx c 2^{a \, t}$ which is the law of binary tree. So, the law we have considered in this analysis is valid \pa{ for not too long ({\it i.e.} $t-t_0<15$h) but not too short time ({\it i.e.} $t-t_0>3$h). } 
\subsubsection*{Space scale}
\begin{figure}[htbp]
\centering
\includegraphics[width=0.45\textwidth]{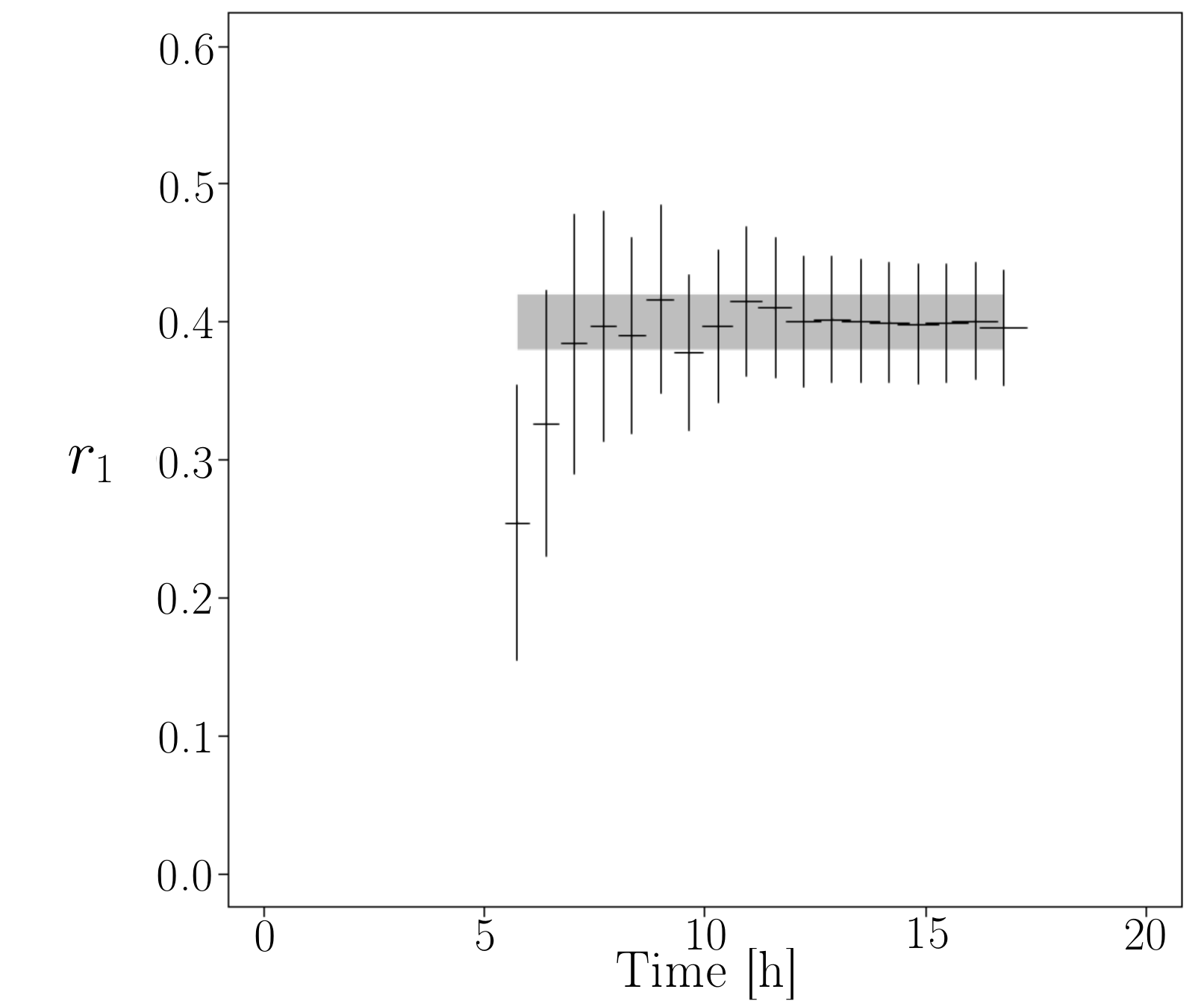}
\includegraphics[width=0.45\textwidth]{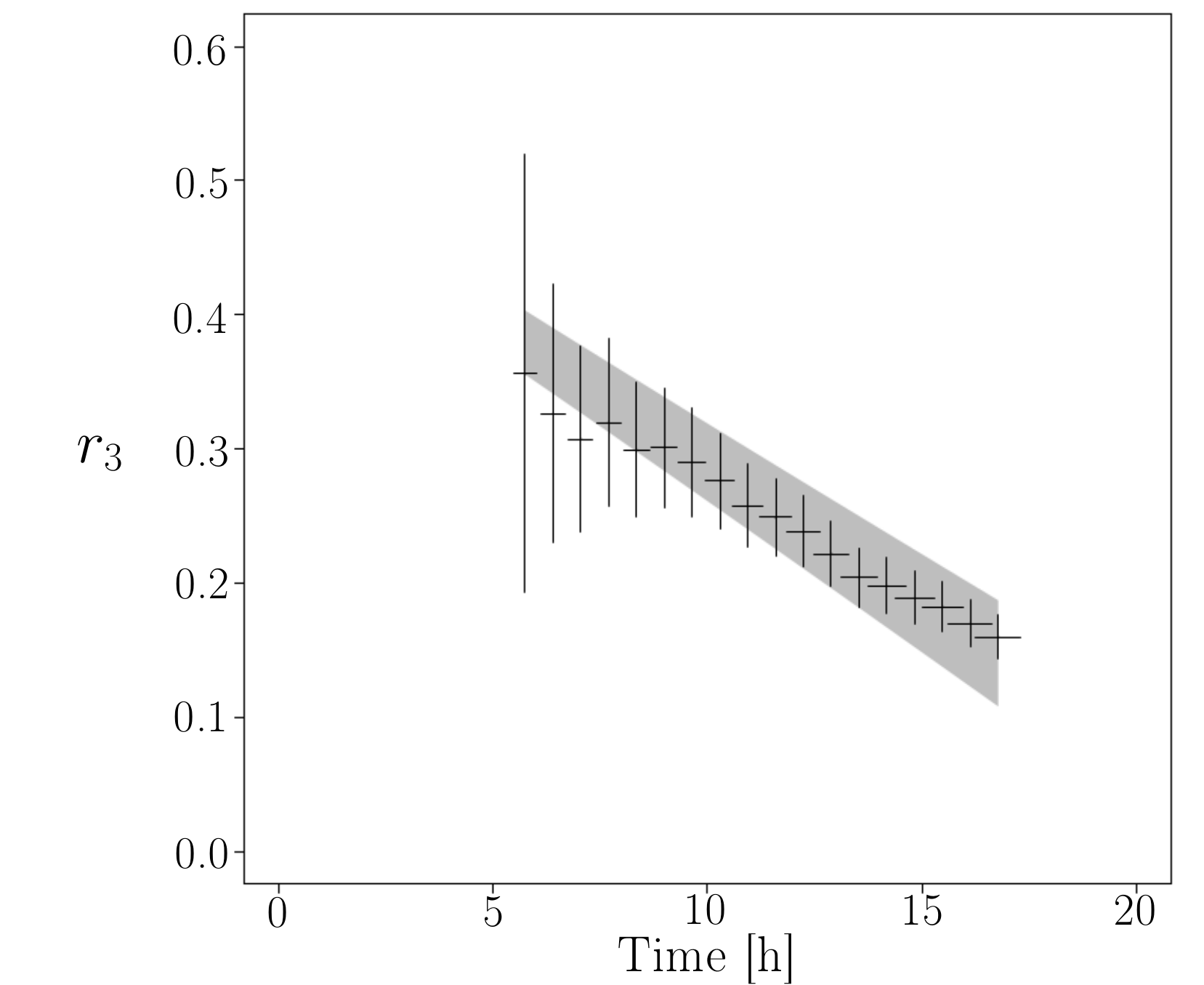}
\caption{(Left) 1-standard deviation range of $r_1^{data}$ experimental data fit in function of time (grey). Black points are $r_1^{sim}$ extracted from the simulation with scaled time ans scaled space. (Right) 1-standard deviation range of $r_3^{data}$ experimental data fit (grey). Black points are $r_3^{sim}$ extracted from the simulation with scaled time and space coming from $V_{1ob}$. See text for details. }
\label{fig4e}
\end{figure}

With a growth model in the form of a binary tree, it is easy to obtain an observable which makes it possible to calibrate the distances in the simulation. For a given generation $g$, the number of vertices $V_1$ is $N_{V_1}=3\,(2^{g-2})$ where the factor three comes from the number of branches of the germination. The number of segments between two vertices (($V_3,V_3$ or ($V_3,V_1$)) is $N_{seg} =3\,(2^{g-1}-1)$, so that the average length of the network can be written $\langle L \rangle=\langle \ell \rangle \, N_{seg}$ where $\langle \ell \rangle$ is the average length between two vertices. The ratio, $r_1$, of the average length of the network to the number of $V_1$ in the network is then $ r_1=\langle \ell \rangle \, (2-2^{2-g})$ so that this is constant for $g\gg1$. \\ 
Because of lateral branches we need to correct this ratio. By construction, the number of lateral branches is proportional to the number of vertices segments. So, $r_1 \simeq \langle \ell \rangle \, (2-2^{2-g}) \, \frac{1+p}{1+2p}$ for $g\gg 1$ and where $p$ is the mean fraction of lateral branches for a branch. $r_1$ is  again a constant for $g\gg 1$. The index of generation $g$ is linked to time that was calibrated previously. \\
So, for a time not to small we must show on data a constant for $r_1^{data}$. This is the case and we can estimate this constant on the data so as to calibrate the species on the simulation. \\
In the Fig.~\ref{fig4e} (left) we plot the uncertainties for one standard deviation giving from the fit parameters of data (grey area). We scale $r_1^{sim}$ by the relation $r_1^{sim}=r_1^{data}$ for a time not to small (using the scaled time obtain before) and plot the ratio coming from simulation (black cross). The uncertainties on the numbers of $r_1^{sim}$ are obtain by considering a Poisson hypothesis and relative uncertainties for $L_{tot}$ equal $5\%$. \\
In Fig.~\ref{fig4e} (right) we show the ratio $r_3^{data}$ coming from the fit (grey area is one standard deviation) and we plot the $r_3^{sim}$ for the simulation with the space scale factor and the time scale factor obtained by $v_{1ob}$. Note that it is impossible to define the law of $r_3^{data/sim}$ because of the geometric intersections. However we know that this ratio must decrease as a function of time since the number of geometric intersections increases strongly and independently of the total length of the network. The agreement for $r_3$ between experimental data and simulation is convincing.

\begin{figure}[htbp]
\centering
\includegraphics[width=0.45\textwidth]{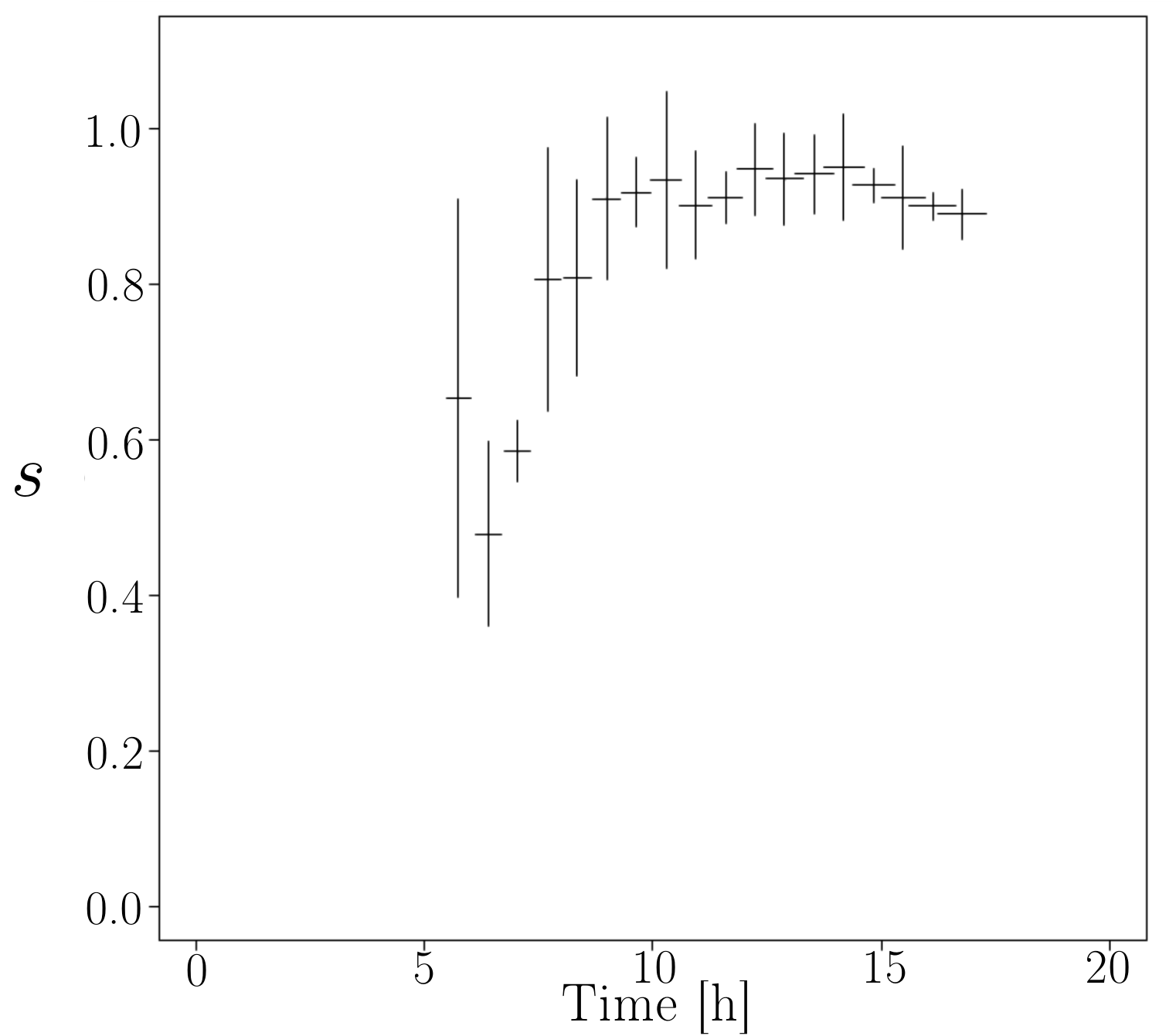}
\includegraphics[width=0.45\textwidth]{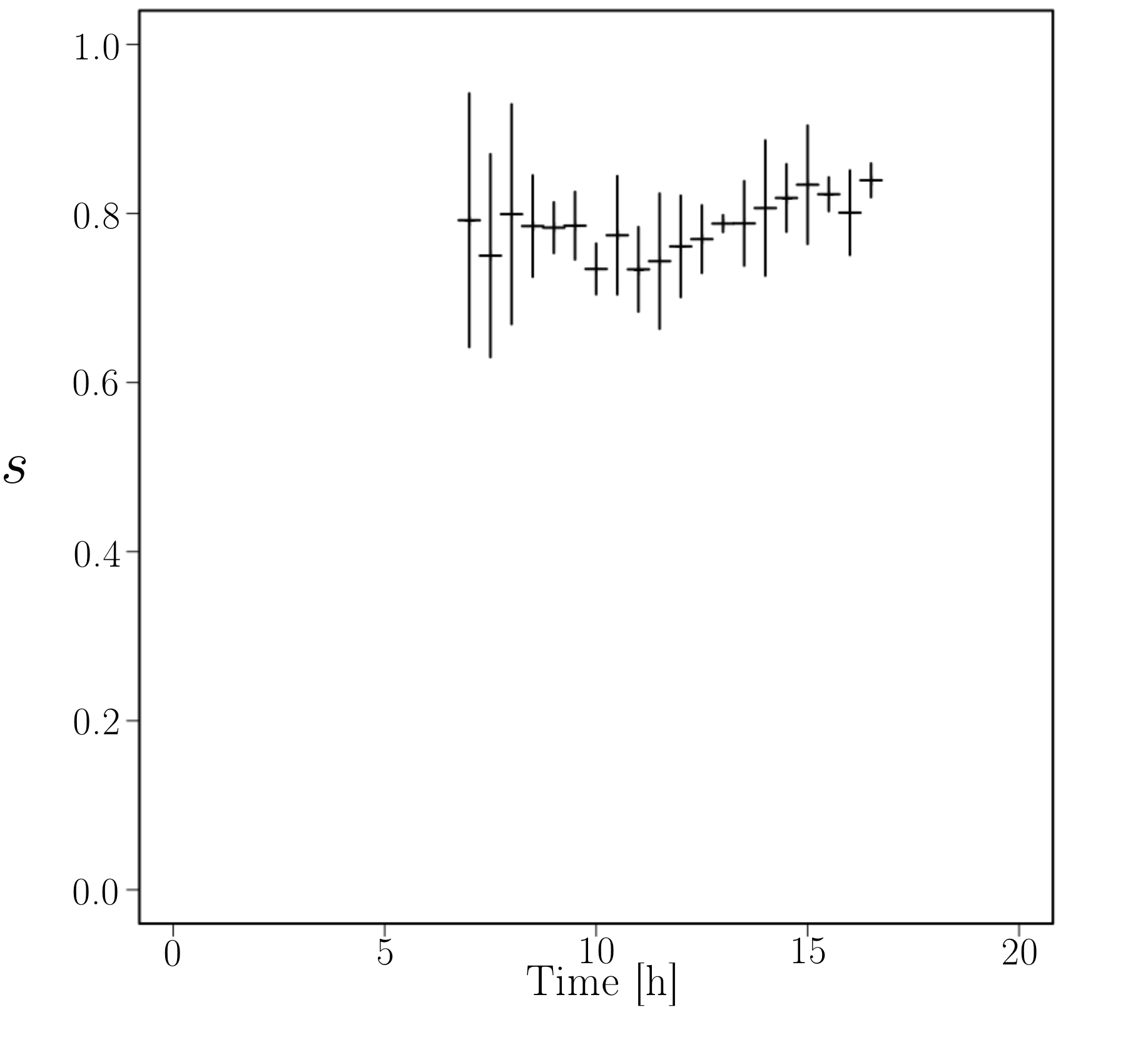}
\caption{Sphericity of $V_{1ob}$ vertices distribution, defined as the ratio $s=2\lambda_2/(\lambda_1+\lambda_2)$ of the eigenvalues (see the text for more details) in function of time. Right, sphericity extracted from the simulation, with scaled time and space. Left, is sphericity extracted from experimental data.}
\label{fig4d}
\end{figure}

\noindent
Another test is to check if the spatial geometry of the network has the same geometric behaviour between the experimental data and the simulation as a function of time. For this purpose, at a given time, we construct the inertia tensor of the vertex distribution for $V_{3ob}$ (resp. $V_{1ob}$):

\begin{align}
    I&=
    \begin{vmatrix} 
        \sum_n (x_n -x_0) \, (x_n -x_0) & \sum_n (x_n -x_0) \, (y_n -y_0) \\ 
        \sum_n (x_n -x_0) \, (y_n -y_0) & \sum_n (y_n -y_0) \, (y_n -y_0)
    \end{vmatrix}
\end{align}

with $x_0, y_0$ the mean position of the vertices cloud and $x_n,y_n$ the position of the vertices. \\
Diagonalisation of this tensor gives two eigenvalues (and two eigenvectors, which are the main axes of the vertices cloud). These eigenvalues are ordered as $\lambda_1 \geq \lambda_2$ and we define an indicator of the geometry of the vertex cloud: $s=\frac{2\lambda_2}{\lambda_1 + \lambda_2}$ (called {\it sphericity}). If $\lambda_1 \gg \lambda_2$ then $s \rightarrow 0$, which means that the vertices cloud has the shape of a \pa{elipse}, if the eigenvalues are degenerated, $s=1$, which means that the vertices cloud has the shape of a \pa{disk}. \\
In the data the shape of the vertex cloud is concerted over time ($s(t)=constant$). 
Its value depends on the initial conditions, {\it i.e.} the way the spore was generated. So the algebraic value of the constant is not relevant and only the shape of $s(t)$ is relevant. \\
We check if the simulation gives a constant sphericity with scaled time and space. Fig.~\ref{fig4d} gives $s(t)$ for the simulation and for the data of the $V_{1ob}$ distribution in the plan. The values for the uncertainties are obtained by a bootstrap method. \cite{b_efron_introduction_nodate} The sphericity for the $V_{3ob}$ vertices are also constant for data and simulation but because of the $V_{3i}$ it is difficult to give a right interpretation. \\
We check whether the direction of the main axes of the vertices cloud move as a function of time. In the case of experimental data this direction is constant even $s\sim 1$. The same applies in the case of the simulation. 
\section*{Prediction of angle distribution}
\begin{figure}[htbp]
\centering
\includegraphics[width=0.5\textwidth]{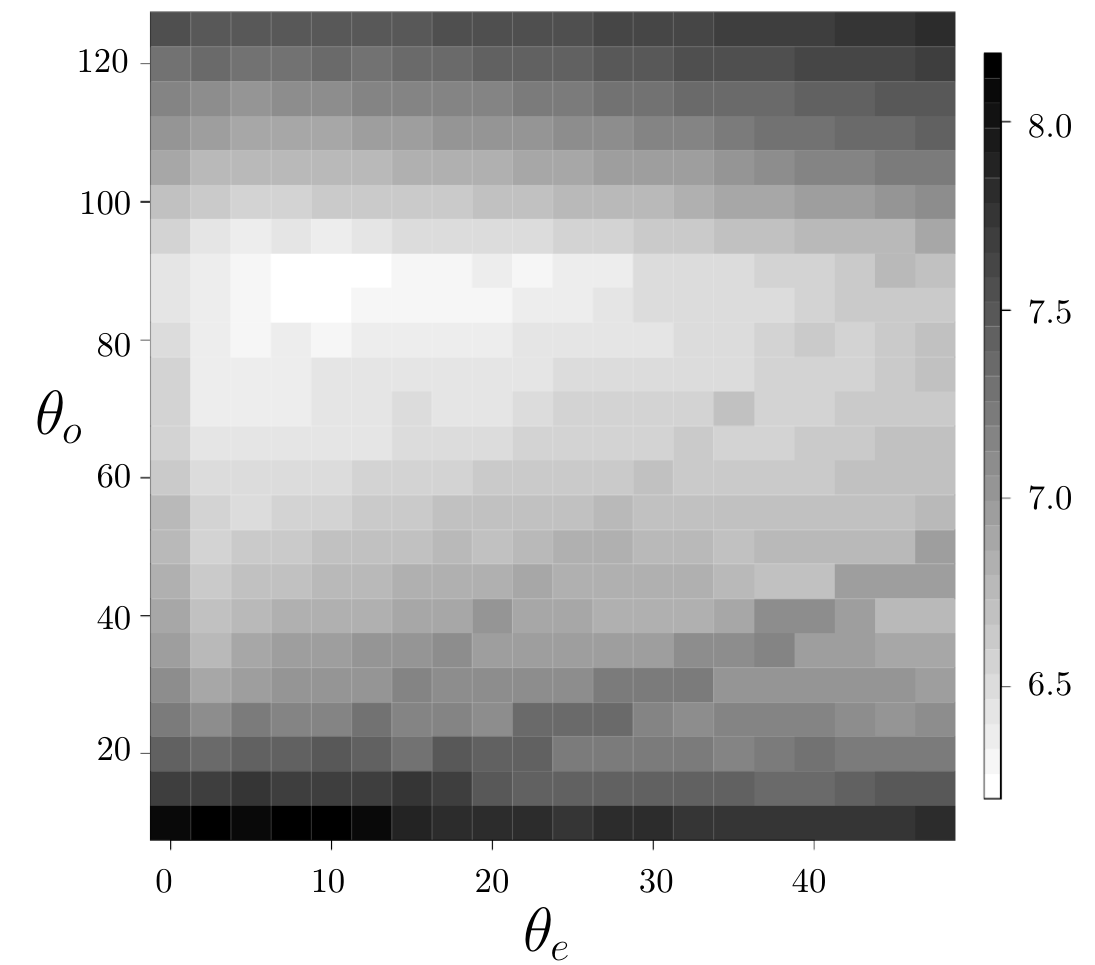}
\caption{Logarithmic distribution ($z$-direction) of the number of times the branches intersects each other in the network as a function of branch angles $\theta_o$ ($y$-direction) and $\theta_e$ ($x$-direction) in degree for a fixed time and length.}
\label{fig6}
\end{figure}

As noted above, when a $V_1$ vertex is mutated to a $V_3$ vertex (an apical branching) the two ``daughter'' branches have very distinct angles. We can make the hypothesis that the permanence of this behaviour is related to very deep reasons. There are undoubtedly biological reasons, but we can hypothesize that the source of this phenomenon is due to the fact that the fungus in a homogeneous environment has a clear advantage in occupying the largest surface in order to capture the maximum of resources from the environment and thus optimizes its growth. The surface occupied by the fungus can be written : 

\begin{equation}
    S_{occ}(t,L,\theta_e,\theta_o) = S_{bio}(t,L) - N_{V_{3i}}(t,L,\theta_o,\theta_e) \Delta S(t,\theta_o,\theta_e)
\end{equation}

with $L$ the total length of the network, $S_{bio}$ the physical surface, or total area, of the fungus and $\Delta S $ the surface of the intersections overlap (because the branches are flat tubes with constant width). \\
If we fix the time and the total length of the network, the surface occupied by the fungus is all the more important as the number of intersections is small for an overlapping surface. It is possible thanks to the simulation to test this hypothesis in order i) to verify that indeed the largest occupied surface requires two quite distinct angles, and ii) to predict the optimal couple of angles. Fig.~\ref{fig6} shows the logarithm of $N_{V_{3i}}(t,L,\theta_o,\theta_e)\Delta S(t,\theta_o,\theta_e)$ as a function of $\theta_o$ and $\theta_e$ for a fixed time ({\it i.e.} generation index). Building upon the previous assumptions, we can make the following predictions, the wide angle is close to $80^{\circ}$ and the small angle is close to 15$^{\circ}$. The uncertainties on theses angles are estimated to be about 10$^{\circ}$ because of the toy model used here (the fixed length of the branches at a fixed time about nine generations).  \\

\section*{Experimental approach}

{\it Podospora anserina} is a coprophilous filamentous ascomycete, a large group of saprotrophic fungi that mostly grows on herbivorous animal dungs and plays an essential role within this complex biotope in decomposing and recycling nutrients from animal feces. \cite{sarrocco_dung-inhabiting_2016} {\it P. anserina} has long been used as an efficient laboratory model to study various biological phenomena, especially because it rapidly grows at a rate of $~7$\,mm/day on standard medium, it accomplishes its complete life cycle in only one week, leading to the production of ascospores, and it is easily usable in molecular genetics, cellular biology and cytology. \cite{silar_podospora_2020}

\subsection*{Experimental set-up}
\noindent
An experimental device allowing to solve the dynamics of the local and global growth of the complete hyphal network of {\it P. anserina}  directly on a Petri dish from an ascospore and over a period of $\sim 15$ hours in a controlled environment has been previously developed and described~\cite{dikec_hyphal_2020}.  
We did use of this setup to carry out three complete and independent series of images of the thallus growth, named $(1)$, $(2)$ and $(3)$ thereafter, under standard growth conditions with M2 culture medium, see \cite{podospora}  for more details, at a temperature of 27$^{\circ}$C.

\begin{figure}[htbp]
\centering
\includegraphics[width=0.9\textwidth]{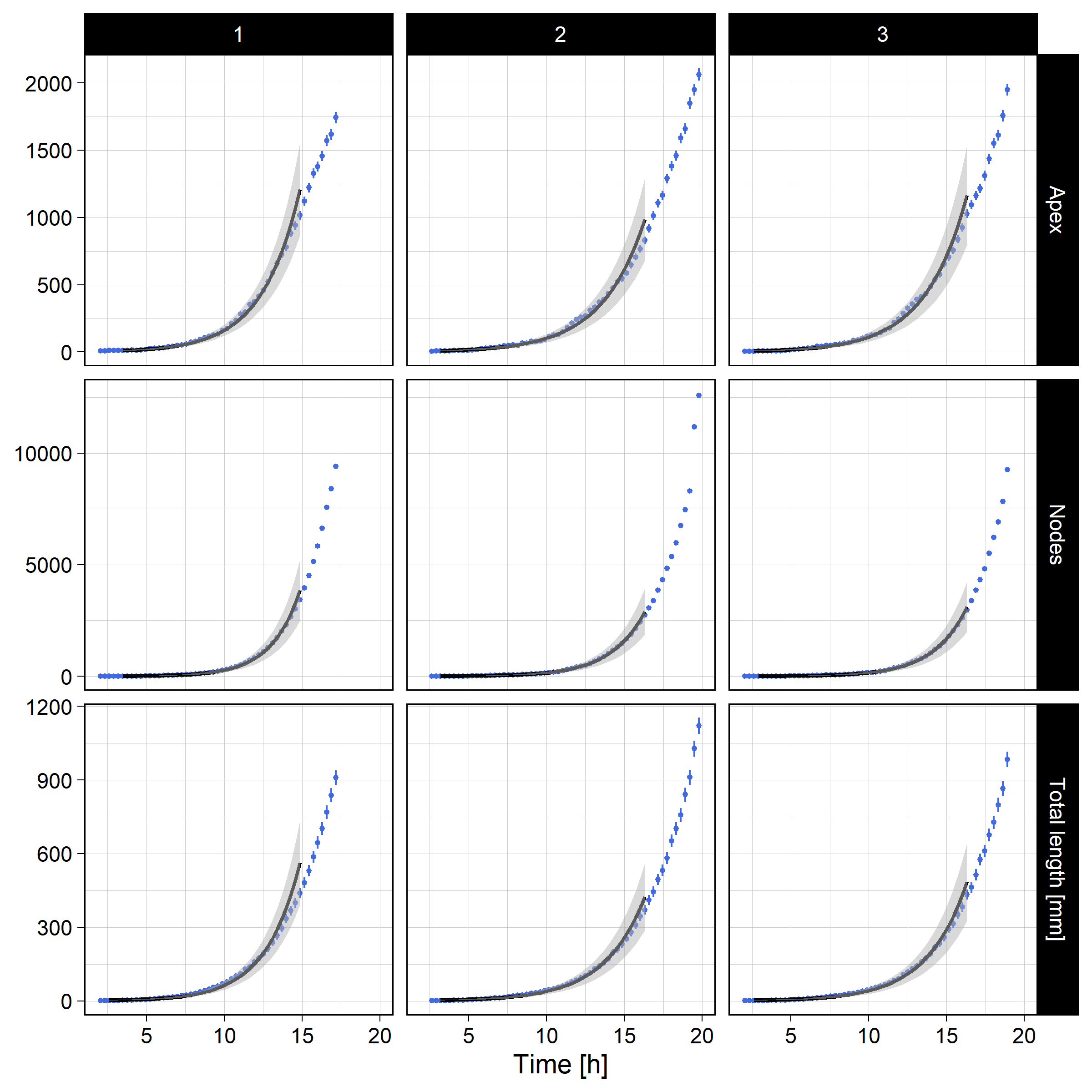}
\caption{Number of apexes $N_{V_{1ob}}$, Number of nodes $N_{V_{3ob}}$ and total length $L$ as a function of time $t$ for the experiments $(1)$ (left), $(2)$ (middle) and $(3)$ (right). Experimental data (blue) are shifted in time of respective $t_0$.
Solid black lines represent the best fit parameters (see Table~\ref{tab:exponent}). The respective grey shadowing wraps the data range used to perform to statistical fit, while its thickness quantifies the associated uncertainties to one standard deviation.}
\label{fig:ANL}
\end{figure}

\noindent
After the standard binarisation and vectorization process described in~\cite{dikec_hyphal_2020} we extracted and adjusted the following quantities in order to calibrate the simulation process. Let $N_{V_{1ob}}$ be the number of observed $V_{1ob}$, $N_{V_{3ob}}$ be the number of observed $V_{3ob}$ and $L$ the total length of the network. We assume the uncertainties associated to the vertices count to be Poisson. The total length $L$ is about $N_{V_{1ob}}+N_{V_{3ob}}$ segments of average length $\langle \ell \rangle$. An estimation of the uncertainty associated with the measure of the total length $L$ was derived as $\sigma_L = \sqrt{N_{V_{1ob}}+N_{V_{3ob}}}\langle \ell \rangle$. We assume the uncertainty associated to the acquisition time $t$ is half the sampling period $\delta t$. 
The growth of such a microorganism follows basically a five steps timeline: i) a lag phase, ii) an exponential phase, iii) a deceleration phase, iv) a stationary phase, and v) a decline phase. We study only the tree first steps. We previously showed that the temporal growth of $N_{V_{1ob}}$, $L$ and $N_{V_{3ob}}$, to a first approximation, can be modelled by an exponential growth. This of course can only be valid  locally (\textit{i.e.} for short growth times) and from a purely descriptive perspective. Indeed, to be consistent with the binary tree model we made use of a base-2 exponential function to describe the growth of these quantities. 
Thus the following expression may be used: 

\begin{align}
    X_i(t) &= X_i^0\;  2^{(t+t_0)/\tau_i}
    \label{eq:fitALN}
\end{align}

\noindent
where $X_i$ stands respectively for $N_{V_{1ob}}$, $N_{V_{3ob}}$ or $L$ and $t$ is the time. $\tau$ is the characteristic growth time, $t_0$ is the temporal offset corresponding to the transition between the end of the lag phase and the first observation. For the experimental data, there is necessarily a time offset between the first recorded image and the ascospore germination. In order to compare the experiments with one another, this time offset should be taken on a per experiment basis because it varies according to each experiment.  $X_i^0$ are the respective parameters $N_{V_{1ob}}^0$,  $N_{V_{3ob}}^0$ and $L^0$, corresponding to the growth onset given the temporal offset. $N_{V_{1ob}}^0$, $N_{V_{3ob}}^0$ and $L^0$ values are respectively expected to be approximately 3, 1, and in the range 10 to 20 hyphal diameters. 
The equation~\ref{eq:fitALN}  shows a non-linear behaviour of the fit parameters. Consequently we made use of the following procedure: 
We excluded spurious data, {\it i.e.} for time greater than $15$\,h and for time smaller to $~2$h. The value of $t_0$ was manually adjusted in order to obtain simultaneously $N_{V_{1ob}}^0$, $N_{V_{3ob}}^0$ and $L^0$ in the range of the respective expected values. We assume that the uncertainty on $t_0$ corresponds to $2 \delta t$. Note that $t_0$ gives access to the lag phase duration.
With $t_0$ as a fixed parameter the least squares fit is linear. The grayed areas in Fig.~\ref{fig:ANL} show the results of the fit (for one standard deviation) on the considered data range. The points are the data with their uncertainties. \\ 
The growth rate exponents $\tau$ and the measurement of specific quantitative parameters for the three experiments are summarized in Tab.~\ref{tab:exponent} and Fig.~\ref{fig:ANL}, respectively.  
As expected with the exponential behaviour shown in Eq.~\ref{eq:fitALN} the algebraic values of the correlation  $\rho(\tau,X_0)$ is about $~0.8$. The $\chi^2$ statistics of the fits are found to lie in the range 1 to 5 for a d.o.f about $30$. 
\begin{table}[htbp]
\centering
\begin{tabular}{|c|ccccccc|}
\hline
    & $t_0$   
        & $N_{V_{1ob}}^0$  
            & $\tau_{V_{1ob}}$  
                & $N_{V_{3ob}}^0$  
                    & $\tau_{V_{3ob}}$   
                        & $L^0$   
                            & $\tau_L$ 
                                \\
\multirow{-2}{*}{Experiment}    
    & {[}h{]}    
        & {}   
            & {[}h{]}     
                & {}    
                    & {[}h{]}    
                        &{[}mm{]}  
                            & {[}h{]}   
                                \\ \hline
$(1)$
    & 1.7  
        & $2.54\pm0.57$   
            & $1.67\pm 0.08$   
                & $1.23\pm 0.37$   
                    & $1.28\pm 0.05$  
                        & $0.81\pm 0.28$   
                            & $1.59\pm 0.09$ 
                                 \\ 
$(2)$
    & 2.3   
        & $2.72\pm 0.55$   
            & $1.92\pm 0.09$  
                & $1.63\pm 0.46$  
                    & $1.51\pm 0.06$ 
                        &  $0.91\pm0.31$   
                            & $1.85\pm 0.11$  
                                 \\
$(3)$
    & 1.7  
        & $2.54\pm 0.52$   
            & $1.84\pm 0.09$   
                & $1.47\pm0.43$   
                    & $1.47\pm 0.06$   
                        &  $0.84\pm 0.29$  
                            &  $1.79\pm 0.11$ 
                                 \\ \hline
\end{tabular}
\caption{Summary of the growth rate exponents for $N_{V_{1ob}}$, $N_{V_{3ob}}$  and the total length $L$ extracted from data shown in Fig.~\ref{fig:ANL}. The $\chi^2$ values were all found to fall into the range 1 to 5.}
\label{tab:exponent}
\end{table}
\\
\noindent
Because these 3 measurements are i.i.d. we assume the final results for the pertinent parameters of the growth are those shown in Tab.\ref{tab:angle_final}.  Note the difference between the doubling time $\tau_{V_{1ob}}$ and  $\tau_{V_{3ob}}$. This is a straightforward consequence of the contribution of the geometric vertices in the 3-body vertices collection .
Conversely, doubling time for the 1-body vertices $\tau_{V_{1ob}}$  and total length $\tau_{V_L}$ are found to be in agreement. 

\begin{table}[htbp]
\centering
\begin{tabular}{|ccc|}
\hline
$\tau_{V_{1ob}} $ [h]  
    & $\tau_{V_{3ob}} $ [h]  
        & $\tau_L $ [h]  \\ \hline
$1.81\pm 0.05$ 
    & $1.42\pm 0.03$ 
        & $1.75\pm 0.06$ \\
\hline
\end{tabular}
\caption{Final results of the experimental growth rate exponents, from data shown in Tab.\,\ref{tab:exponent}}
\label{tab:angle_final}
\end{table}

\subsection*{Direct measurement of angles}
Angles $\theta_o$ and $\theta_e$ formed by an apical branching (see Fig.~\ref{fig:schema} and Fig.~\ref{fig:method}) were extracted using direct reading on the network picture.
At this stage, the thallus network is in its exponential growth phase, with approximately 1500 apexes and a length of 600~mm (Fig.~\ref{fig:ANL}). 
We focused our analysis on apical branching. By observing the dynamic branching process real $V_{3}$ vertices can be segregated from the geometric vertices $V_{3i}$ due to overlapping intersections. The comparison between images acquired before and after the branching allows the operator to extract a real $V_{3}$ vertices collection without any ambiguity. 
The procedure of angles extraction is shown in Fig.~\ref{fig:method}. A circle with a radius of 50~pixels (approximately 5 hyphal diameters) is centered on each $V_{3}$ vertex. 
The coordinates of the points of intersection between the circle and the three hyphae are manually recorded. For each apical branching, the extension of the mother hypha in the direction of growth determines two angles with both emergent hyphae, a small angle and a wide angle. From the complete network, 66 $V_{3}$ were extracted from each experiment. We then considered two populations, respectively composed of the collection of the small angles ${\theta_e}$ and of the wide angles ${\theta_o}$ (Figure~\ref{fig:method}). 
The uncertainties on each angle value $\delta\theta=4$~\text{$^\circ$} is estimated from  the diameter of the hyphae.

\begin{figure}[htbp]
\centering
\includegraphics[width=0.45\textwidth]{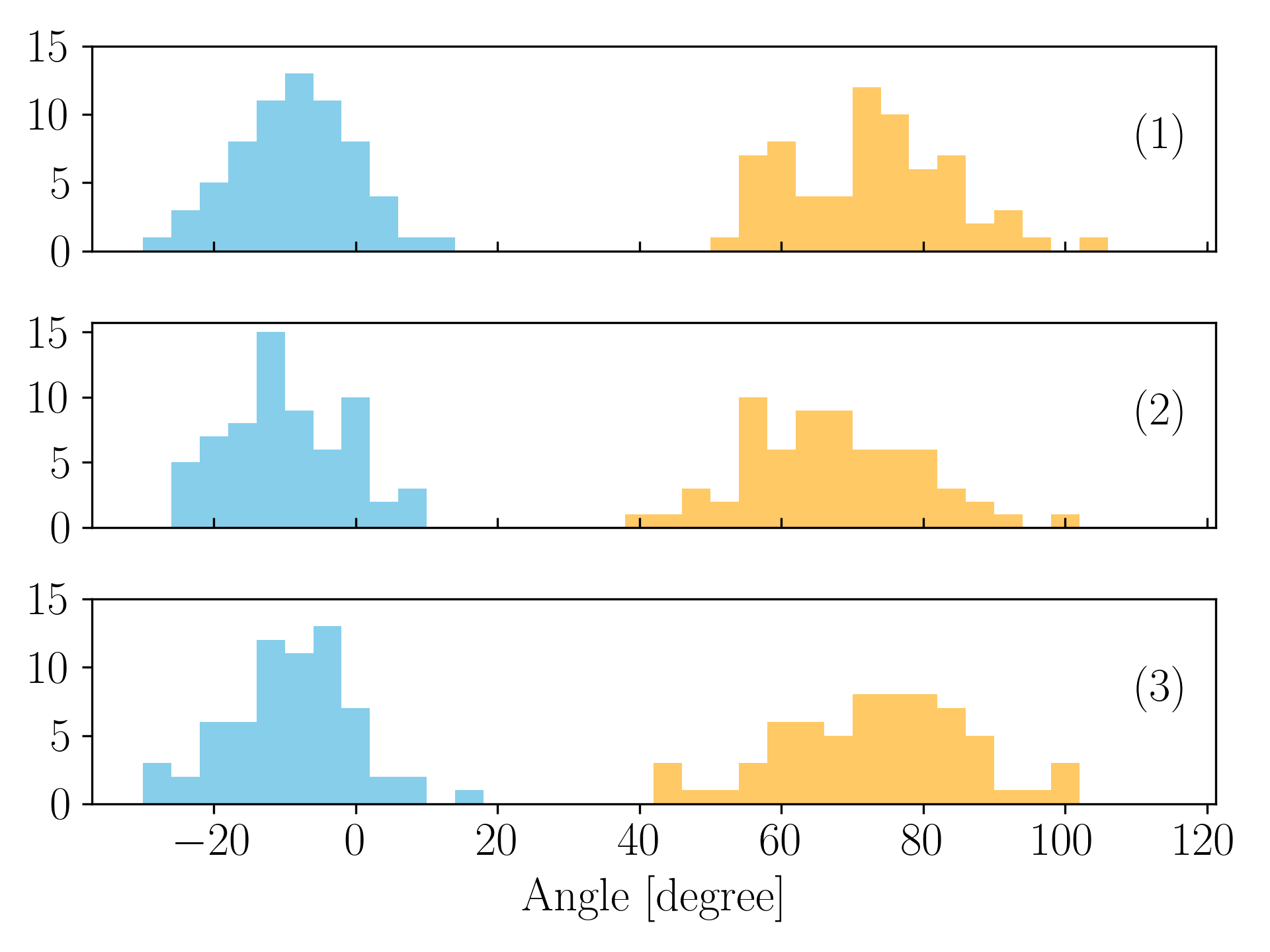}
\includegraphics[width=0.45\textwidth]{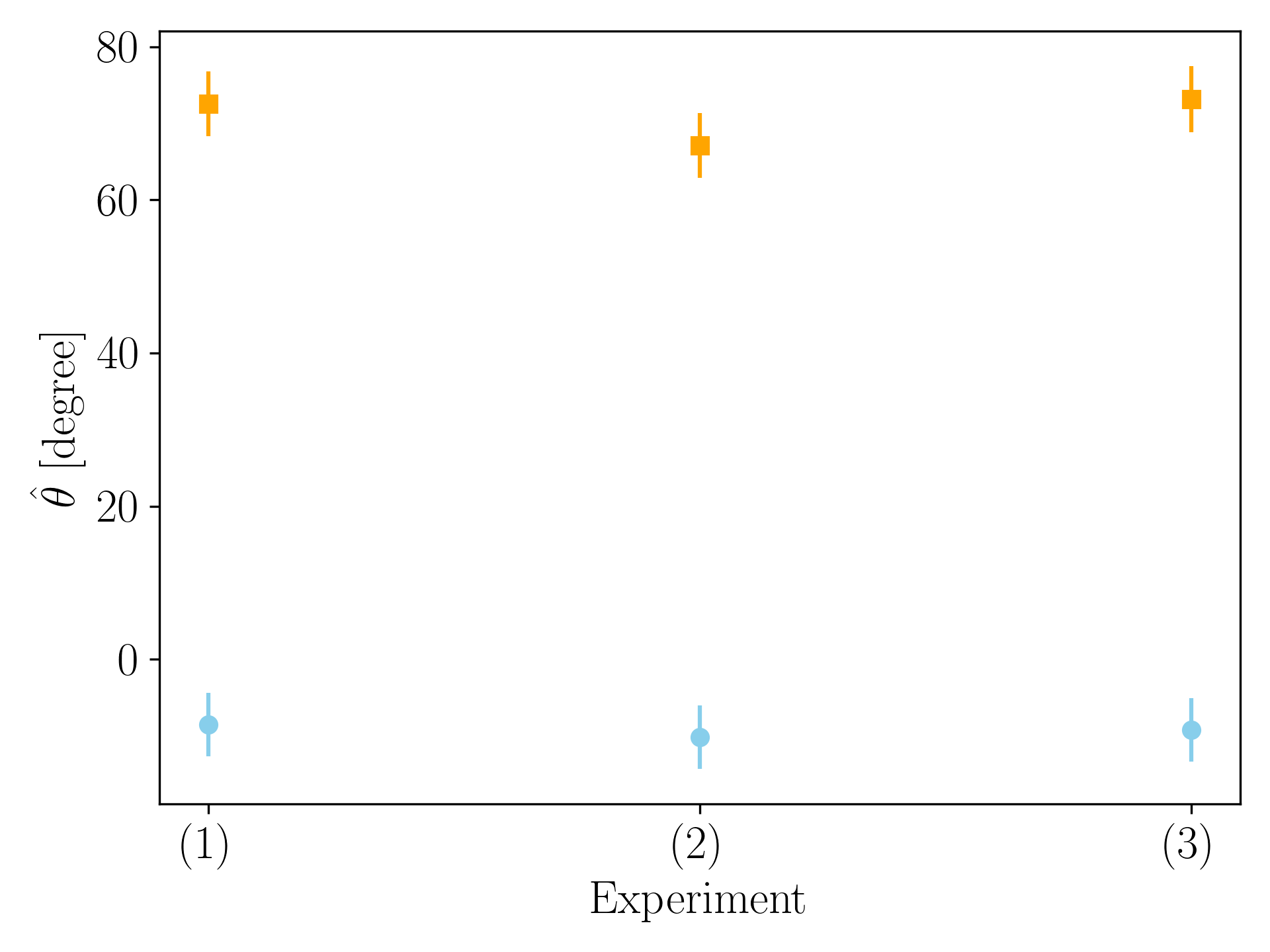}
\caption{(Left) $(1)$, $(2)$ and $(3)$ are representations (pdf) of the populations of 66 measurements of the small angle $\theta_e$ (in blue) and the wide angle $\theta_o$ (in orange), defined in regard to the extension of the mother hypha, as defined in Fig.~\ref{fig:method}. Bin width is 4 degrees in this representation. 
(Right) Gaussian fits (see the text) allow to derive estimators of the means with uncertainties and widths of the Gaussian (not shown).}
\label{fig:distributions}
\end{figure}

\noindent
Fig.~\ref{fig:distributions} depicts the distributions of the  measurements of ${\theta_e}$ and ${\theta_o}$. Bin width reflects the uncertainties $\delta \theta$. For the clarity of the representation, $\theta_o$ defines the  positive direction of rotation for each $V_{3}$. For each extraction two populations  are clearly visible. Assuming these populations follow Gaussian laws we extracted the respective estimators of the mean $\hat{\theta}$ and width $\hat{\sigma}$ such as $N(\theta_i | theta, \sigma) $ using likelihood maximisation. The uncertainties on angle measurements is assumed to follow a centered Gaussian law with a standard deviation of $\delta \theta$, such as $N(\theta_i | 0, \delta \theta)$. 
The respective estimators are shown in Fig.~\ref{fig:distributions} (right). The corresponding covariance matrices are assumed to have $cov(\hat{\theta},\hat{\sigma})=0$ because the correlation between the estimators are very tiny. We finally compute an  average estimate from this collection of i.i.d sets for each of the two angles, $|\hat{\theta_o}|=70.9\pm 2.4$ and $\hat{\sigma_o}=12.0\pm 2.4$, $|\hat{\theta_e}|=9.3\pm 2.4$, $\hat{\sigma_e}=8.7\pm 2.4$. \\
Since we must take into account the systematic uncertainties $\delta \theta = 4^0$, the final results for the angles are :  $\hat{\theta_o}=71\pm 5$ $\hat{\theta_e}=10\pm 5$.
\begin{figure}[htbp]
\centering
\includegraphics[width=0.45\textwidth]{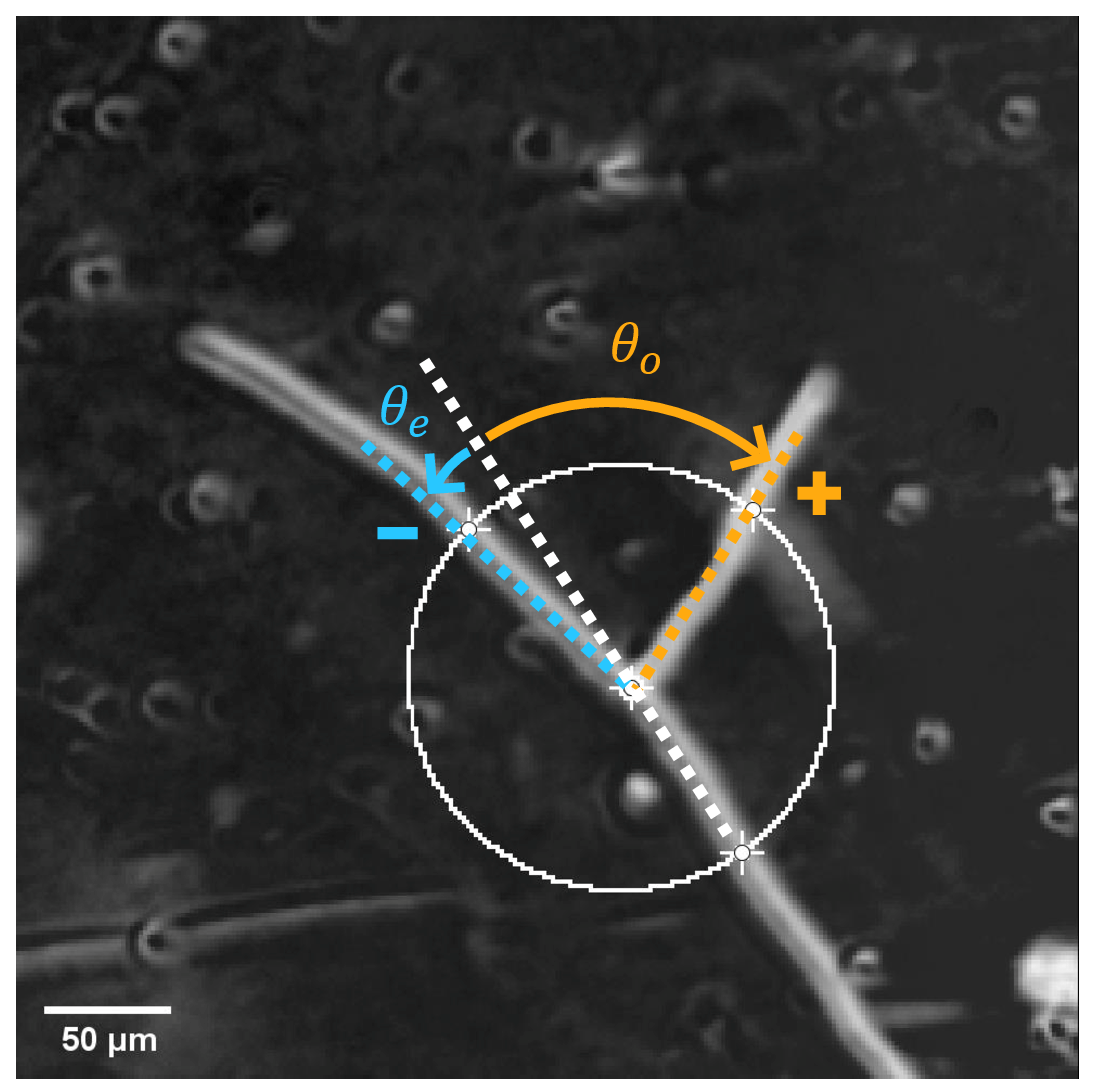}
\caption{
Typical branching process. A first hypha  grows from the bottom right of the image. The branching reveals two new hyphae. A circle of radius $R=80$~$\mu$m~$\sim 5$ hyphal diameters and located on the branching point is drawn. Small angles $\theta_e$ and wide angles $\theta_o$ are defined by the intersections of this circle with the new hyphae and the projection of the first (mother) hypha. Following the convention proposed into the text, $\theta_o >0 $ and in this example $\theta_e < 0 $.}
\label{fig:method}
\end{figure}

\subsection*{GIS automatic method for angle measurements}
In addition to the work on direct measument angles, which focuses on a set of selected apical branching, we have implemented a Geographic Information System (GIS) automatic method for detecting, in a well-developed thallus, all angles for both apical and lateral branching. While being a non-selective method, it nevertheless offers a global information of branching process, with a high number of angle measure extraction.
GIS geoprocessing was performed for the thallus skeletonization (steps $(i-iv)$) and for the generation of vertices (step $(v)$). \\

\subsubsection*{Skeletonization stage and vertexes detection} 
Several methods for generating centerlines, or skeletons, from polygon features have been described in the literature. We can cite Voronoi diagram based method, medial axis transformation algorithm or Delaunay triangulation method \cite{lewandowicz_method_2020, szombara_comparison_2015}. Among those, an automatic Geographic Information System algorithm, ESRI Polygon to Centerline -- available online thanks to the ESRI\cite{GIS} plateform -- appears to be efficient for modelling skeletons of elongated polygonal features \cite{lewandowicz_method_2020}. The \textit{Polygon to Centerline} GIS tool used in this work, based on the creation of Thiessen polygons, is described as follows:
\begin{figure}[htbp]
\centering
\includegraphics[width=0.9\textwidth]{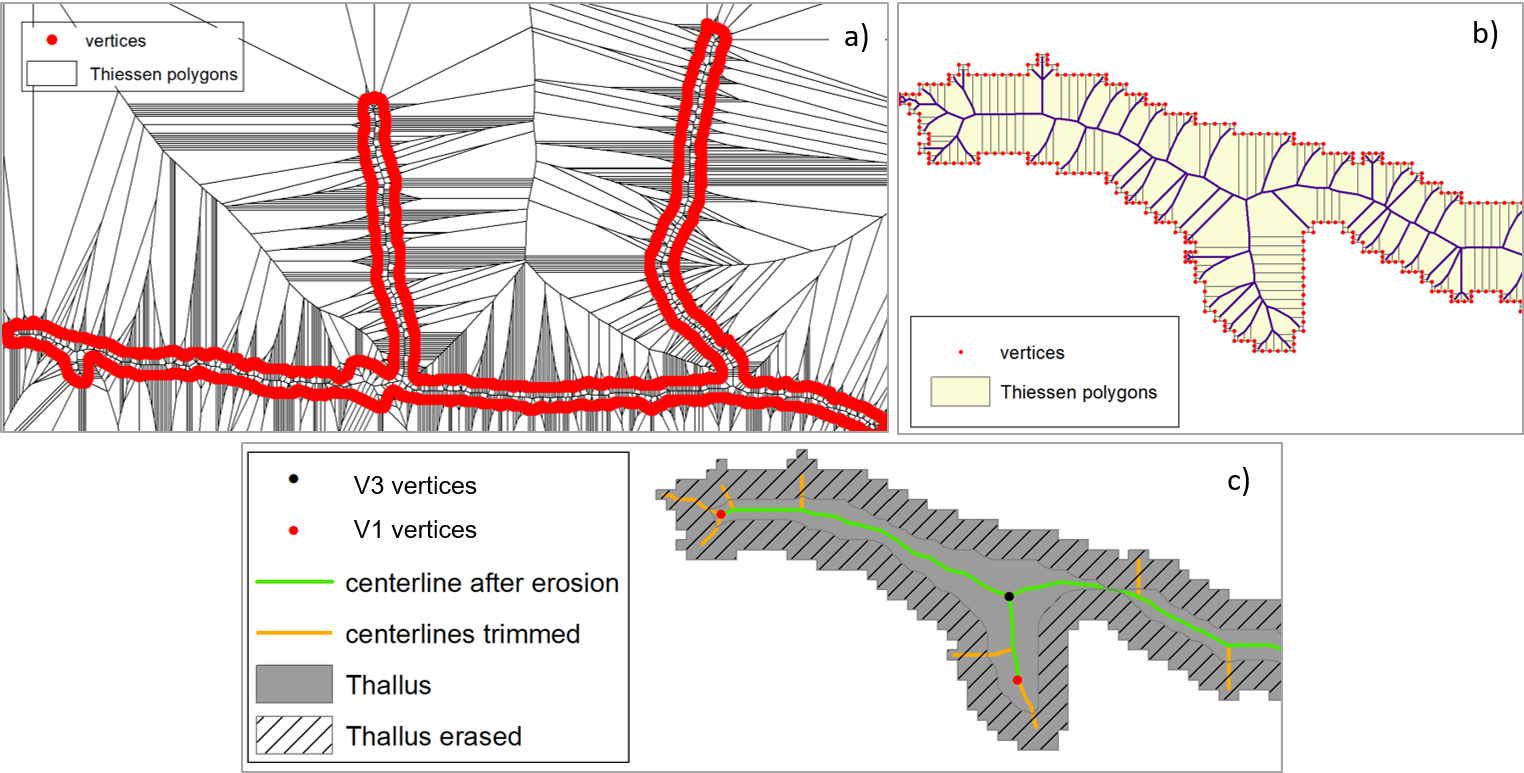}
\caption{Skeletonization processes of the thallus. a) Construction of Thiessen polygons and generation of thallus vertices. b) Zoom in the left part of Fig.\,\ref{fig15}-a shows the point-features vertices of the thallus; Thiesson polygons inside the thallus; the first set of centerlines kept is represented in purple. c) Final geomatic treatments providing the skeleton (green) after the trimming of the short surrounding line segments (the orange ones).}
\label{fig15}
\end{figure}
\\
$(i)$-The first step is to create a large number of vertices from a thallus polygon feature (methodological aspects were discussed in \cite{dikec_hyphal_2020}) through an adaptive densification method for improving the centerline quality. Vertices are then converted into point features (red point features in Fig~\ref{fig15}-a). 
\\
$(ii)$-Then the Thiessen polygons may be created, each of them contains a single point input feature (\textit{i.e.} a vertex). We can observe a high number of Thiessen polygons located both inside and outside the fungal thallus, and of course only Thiessen polygons located inside the thallus area interest us. Thus, we extract polygons that overlay the thallus by using the ``clip'' geoprocessing tool (represented in beige in fig~\ref{fig15}b), which are then converted into linear features. All linear segments that are connected to the thallus single polygon are removed to constitute the centerlines, represented in purple in Fig~\ref{fig15}b. 
\\
$(iii)$-At this step of the treatments, we can see that a high number of short purple linear segments have to be removed (or trimmed). We used the ESRI ``Trim Line'' tool, which consists in removing in our case all purple lines that do not touch another line at both end points. The result of this treatment is shown in Fig.~\ref{fig15}c (the orange plus green segments).  
\\
$(iv)$-A small number of branching artifacts still need to be removed (the orange ones). The hyphal width being relatively homogeneous, orange centerlines are removed by calculating a morpho-mathematic erosion (creation of a 3-pixel 'negative' buffer from the thallus envelope): all the segments that touch the 3-pixel wide surrounding area (grey shaded area in Fig~\ref{fig15}c) are trimmed and only the green segments are kept to form the skeleton. 
It should be specified that this method is efficient to produce centerline features and for working on $V_{3ob}$ vertices, but it slightly alters the position of the apexes ($V_{1ob}$). \\
$(v)$- We then used successively two GIS tools to dissolve all centerline segments into an unique linear feature and to generate a feature class of points for locating the vertices (nodes and apexes). At this processing step, each $V_{3ob}$ vertex is materialized by three superimposed point features when each $V_{1ob}$ vertex corresponds to only one point feature; this makes it possible to distinguish $V_{3ob}$ vertices from $V_{1ob}$ vertices. Finally, $V_{3ob}$ features are dissolved to produce one feature per vertex (instead of three) (see Fig~\ref{fig:cecilia_thalle_vertices}).\\ 
\\
\subsubsection*{Automatic angle measurements} 
\begin{figure}[htbp]
\centering
\includegraphics[width=0.45\textwidth]{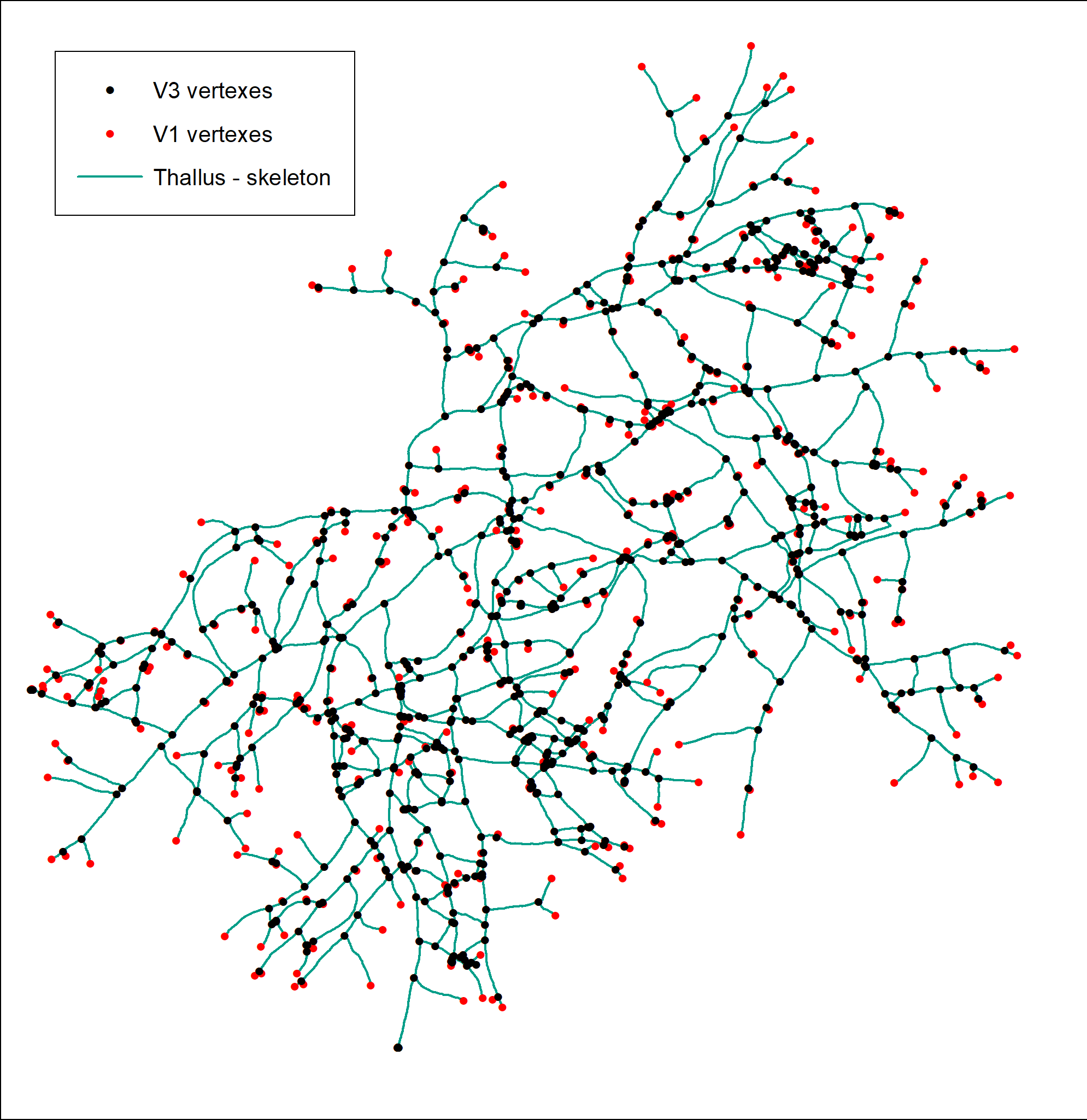}
\includegraphics[width=0.51\textwidth]{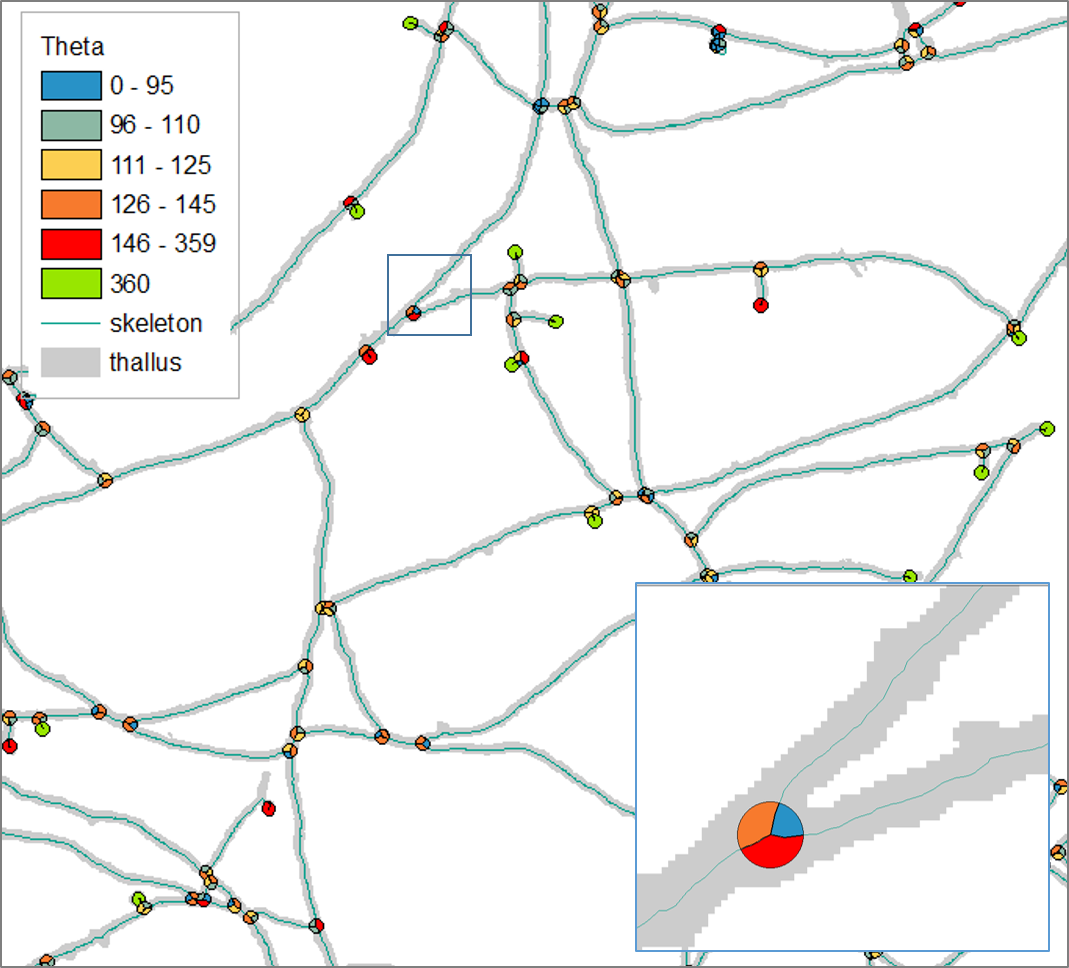}
\caption{(Left) Skeletonization of the thallus, with identification of $V_{1ob}$ and $V_{3ob}$ vertices. (Right) Automatic calculation of angles from 5-pixel buffers around vertices}
\label{fig:cecilia_thalle_vertices}
\end{figure}

The angles were calculated using the following procedure: Discs (or buffers) of 5-pixel radius were centered on apexes ($V_1$) and nodes $V_{3ob}$. All buffers are then splitted by the thallus skeleton. We tested many radius lengths in order to improve the method and we finally settled upon a smaller size radius than those defined for the direct measurement method (different by a factor of 10) due to the high density of line segments (\textit{i.e.} the thallus) in certain portions of the image. Indeed, the latter would have generated additional splits of buffers by non directly connected line segments, which inevitably would have induced a bias in the measurements of disc areas and therefore of angles. The portions of discs are then converted into their equivalent angular counterparts. 
The results we obtained are not significantly different from one method to another. 
Details of this geoprocessing is shown in Fig~\ref{fig:cecilia_thalle_vertices}.
\noindent
A total of 1186 vertices have been detected, with 372 $V_{1ob}$ and 814 $V_{3ob}$, respectively shown by red and black points in Fig.~\ref{fig:cecilia_thalle_vertices}, leading to about 3000 angles.

\begin{figure}[htbp]
\centering
\includegraphics[width=0.45\textwidth]{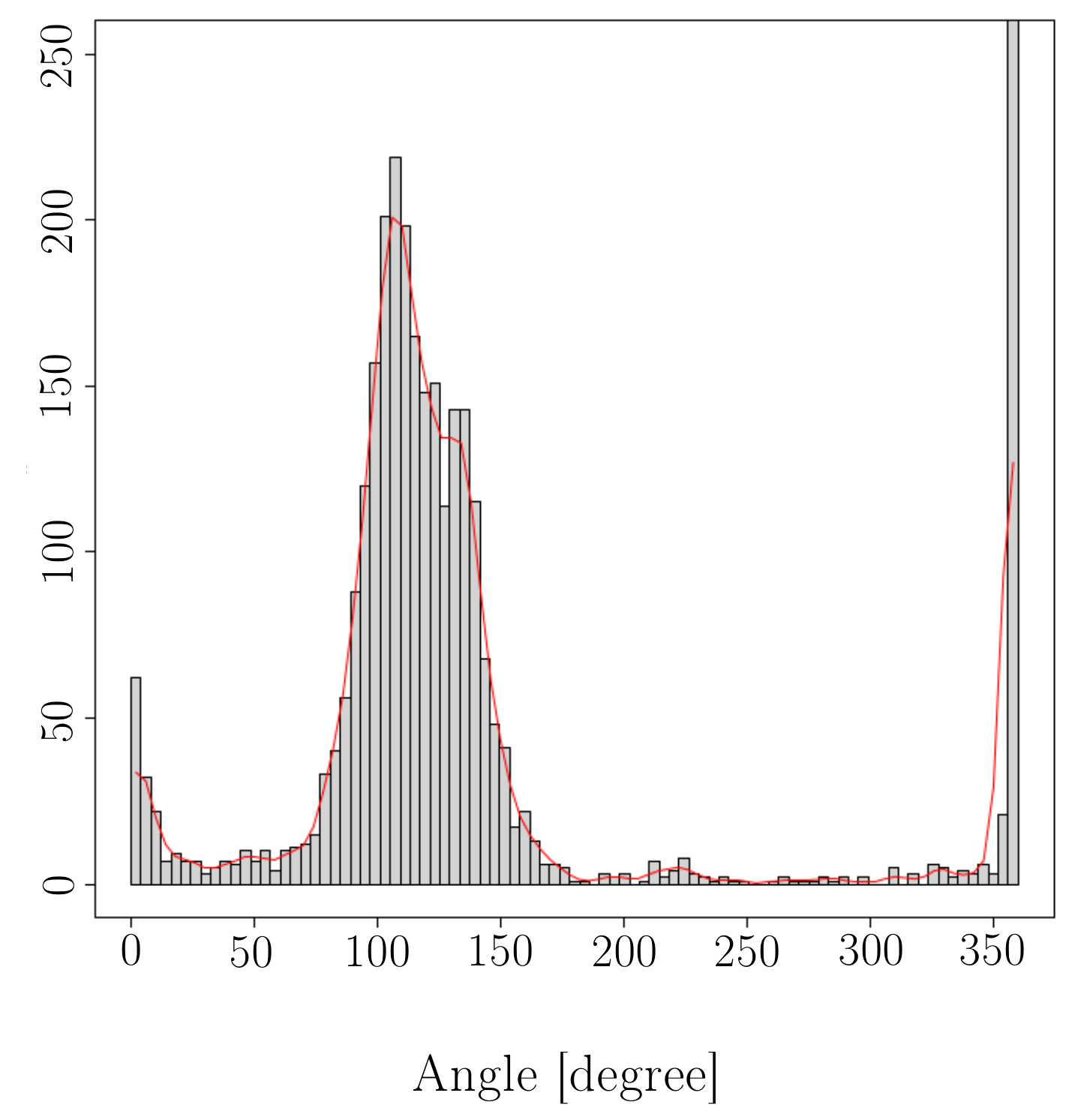}
\includegraphics[width=0.45\textwidth]{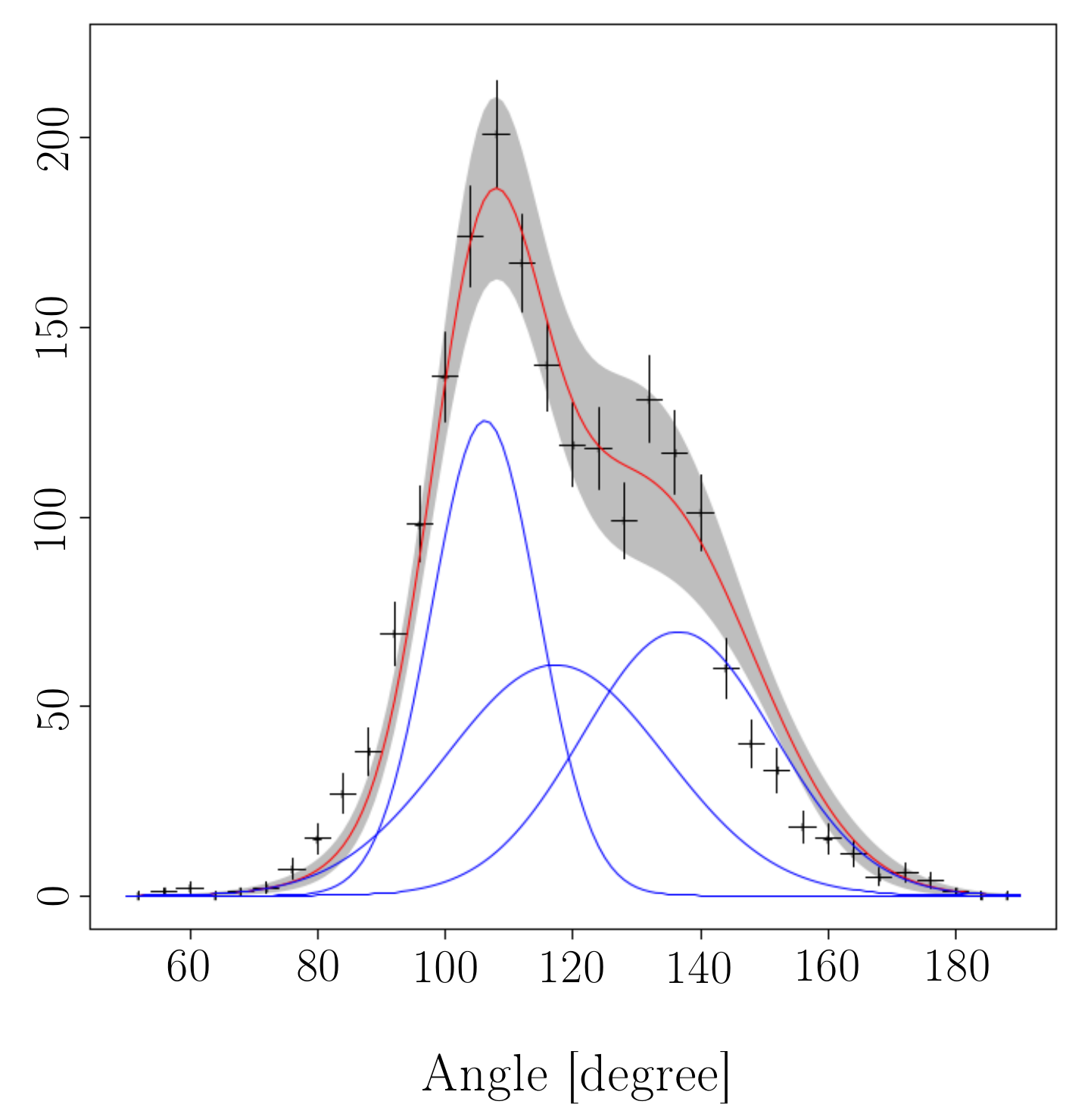}
\caption{(Left) Total angular spectrum with GIS method of the vertices (four populations, three angles from $V_{3ob}$ vertices and one $V_{1ob}$ vertices), as shown in\,Fig.\ref{fig:cecilia_thalle_vertices}. (Right) Zoom in the range of interest. In blue, the three Gaussian distributions. In red the normalised sum of the three Gaussians. In grey, range of 1-standard deviation derived from the fit.}
\label{fig:hist_theta}
\end{figure}

\noindent
Four populations of angles can be distinguished in Fig.~\ref{fig:hist_theta}. The peak observed at 0 degree is an artifact due to a distance less than 10 pixels between two vertices. The peak at 360 degrees corresponds to $V_{1ob}$ vertices that do not partition the circle. Finally peaks at roughly 50 and 200 degrees correspond to the distribution of the 3 arcs of the circle partition. \\

\noindent
This pdf has been fitted using 3 Gaussians (see Fig.\ref{fig:hist_theta}) but one Gaussian is only defined by a normalisation factor (the sum of the 3 angles must be $2\pi$). The $\chi^2$ is very good for about 30 d.o.f. The estimated parameters are respectively $106 \pm 1 $ and $137 \pm 3$ for the means and $8.4 \pm 1$  and  $15 \pm 2$ for the widths of the interest gaussians (the third Gaussian is fully constrained because the sum of the angles must $2\pi$). \\
There is only one geometric scenario to define $\theta_e$ and $\theta_o$ when we have 3 angles and when we know the hierarchy between these angles. The right scenario gives the angle values : $\theta_e = 43\pm 6$ deg and $\theta_o= 75\pm 5$ where the uncertainties take into account the uncertainties on the angle constraint. 
%

\section*{Discussion}

In a previous paper,\cite{dikec_hyphal_2020} we characterized the hyphal network expansion and densification of the filamentous fungus {\it P. anserina} --in particular through the evolution of the total length and the number of nodes and apexes-- on a standard culture medium assumed to be homogeneous and optimal for the fungal growth. We have presented in details the experimental process implemented both from a biological and physical point of view. Here, we present a two-dimensional simulation of the fungal growth that allows us to better characterize some growth patterns of the fungal network. In particular, we predicted that thallus growth is driven by a specific angular branching process, optimized for the exploration and exploitation of the environment.
\noindent
In this work, a calibration of the simulation was first carried out from the collected data. To this end, we focused on the two main processes that determine the growth pattern of the fungal network -\textit{i.e.} apical growth and branching- to propose a growth model in the form of a binary tree, meaning that each branch is divided into two sub-branches, as previously described.\cite{garnier_discrete_2009} Then, in the generation process, the main features of this model are that lengths and angles follow two different probability laws and that growth has to be considered dynamically. The fungal network growth has then been basically defined as an apical branching process characterized by intersection points between mother/daughter branches ($V_3$), and apexes of branches ($V_1$). The growth model was incremented by lateral vertices ($V_{1\ell}$ and $V_{3\ell}$) that can occur on a new daughter branch at a later moment of the growth. Such a process is in agreement with the usual description of fungal network formation.\cite{harris_branching_2008} Namely, two types of branching are allowed to occur according to their location on the hyphae: apical branching which is the emergence of two branches from a hyphal tip (named here $V_3$) and lateral branching from the sub-apical part of one hypha (named here $V_{3\ell}$). The proposed model was calibrated using the parameter values obtained from our experimental data \cite{dikec_hyphal_2020}, namely the number of vertices for different times and the spatial geometry of the fungal network.
\noindent

\begin{table}[htbp]
\centering
\begin{tabular}{|c|cccc|}
\hline
                           &  $\theta_e$   &  $\sigma_{\theta_e}$  &  $\theta_o$   &  $\sigma_{\theta_o}$ \\ \hline  
Simulation (prediction)    & $15\pm 10$  & n.d & $80\pm 10 $ & n.d \\ \hline   
Direct measurement         & $10\pm 5$  & $8.7\pm2.4$ & $71\pm 5$   & $12\pm 2.4$ \\ \hline   
GIS method                 & $43\pm 6$  & $8.4\pm 1$ & $75 \pm 5$  & $15\pm 2$ \\ \hline   
\end{tabular}
    \caption{Results on the angle measurements and prediction}
    \label{tab:angle_resume}
\end{table}
\noindent

\noindent
The quality of experimental observations is limited by geometric intersections (overlapping). Obviously this limitation does not apply to the numerical simulation from which  an estimation of the number of geometric intersections can be derived. Moreover, it unambiguously constructs the objective (and unobserved) history of the development of the thallus which is inscribed in the position of ``real'' $V_3$.
This simulation also allows an exhaustive study of chirality breaks during growth and/or at the origin of development.
\noindent
The most important feature of this model is the occurrence of two distinct angles, a wide angle and a small one at the daughter branches, when a $V_1$ vertex transforms into a $V_3$ vertex. This pattern of angular distributions at the apexes is not the one commonly used by the simulation attempts previously described. For example, \cite{du_lattice-based_2018} has developed a lattice-based system for modelling mycelial growth of {\it Postia placenta} with an apical branching defined as the emergence of two branches that symmetrically develop with respect to the previous tip direction. 
Nevertheless, in our case this typical angular distribution allowed us to describe the apical branching process as the simultaneous emergence of two distinct types of branch. First is an exploratory branch, which slightly deviates (small angle) from the initial trajectory of the mother branch and could thus be involved in the centrifugal exploration of new territory, and then in the extension of the colony. Second is the operating branch, which shows a larger angle and deviates strongly from the initial trajectory of the mother branch. This type of branch could thus contribute to the exploitation of neighboring territories and then to the densification of the network. Such a pattern is in agreement with  previous studies in which a clear distinction was made between leading hyphae at the edge of a fungal colony which grow into new territory in search for food, and the hyphae behind the colony edge that interconnect to form a three-dimensional network optimized to extract nutrients from the surrounding medium.\cite{lew_how_2011} Then, leading hyphae, with an almost-linear trajectory from the center to the front of the thallus could correspond to the iterative development of exploratory branches from the apexes, contributing in turn to the extension of the thallus.  \\
We hypothesized here that in a homogeneous environment, the fungus thallus growth is suited to occupy the largest surface in order to capture the maximum of resources and thus optimize its growth. Based on our hypotheses, our model predict that the largest occupied surface requires two quite distinct angles, a wide angle, close to 80$^\circ$ and a small angle, close to 15$^\circ$ (see Tab.\,\ref{tab:angle_resume}).
In previous studies, the measurement of branch angles has rarely been reported, and it remains in general limited in scope preventing the development of robust statistics. For example, \cite{abd-elsalam_first_2009} showed that morphological characteristics of {\it Rizoctonia solani} included a right-angled branching that could be used in identification of some isolates. Simonin {\it et al.} \cite{simonin_physiological_2012} carried out the measurement of angles on about a dozen leading hyphae of {\it N. crassa} and showed an angular distribution ranging between 50 and 90$^\circ$. More recently, \cite{du_morphological_2016} led a comparative plot of the angular distribution, both for apical and lateral branching on the thallus of {\it P. placenta}. They showed that this distribution is not affected by the age of region, or by the branch type and that branch angles remained approximately at 80$^\circ$, close to the right angle, which appeared to maximize the area covered by the colony. \\
From new sets of experiments, two approaches have been used in order to extract the  branching angular distributions, (see Tab.\,\ref{tab:angle_resume}). One is based on the extraction of the angles formed by an apical branching ($V_3$) using a direct reading on the image, the other one derives from a GIS automatic method that allows for a global detection of angle measurements (apical and lateral branching). The main point of attention for the development of the first approach was the choice of the diameter of the circle allowing to measure the intersections between branches and thus to measure the branch angles themselves. A circle with a diameter of 50\,pixels (approximately 7 hyphal diameters) was retained. This diameter was  chosen in order to decrease the uncertainty due to hyphal width and to decrease the biases due to variation in the orientation of the apexes with time, independently from the initial branching. Branch angles extraction from approximately 200 apical branching allowed us to clearly identified the small and wide angle populations and to numerically estimate their corresponding values. The  agreement is striking  when compared to the optimal angular distributions that emerge from the prediction. However, the main disadvantage of this approach is that it requires the intervention of an operator who must select the apical vertices to be analyzed. This can be tedious, introducing a bias in the choice of apical branching, and necessarily result in a limited number of selected apexes. However, these results are confirmed by a second approach, complementary to the first one, that is global and based on GIS automatized detection of branch angle measurements. Here again, two populations of angles have been detected, with angular distributions that regain correctly the occurrence of a small and wide angles on apical branching. 

Comparison of numerical values leads to several methodological remarks concerning the wide and small angles. Wide angles were found in good agreement with direct measurement and prediction, within one standard deviation for the two angles. Small angles were found significantly higher for the small angle between GIS measurement and direct measurement (and simulation).
Differences observed for the small angle measurements between the direct approach and the GIS approach could have several sources. Firstly, the choice of the vertices of the direct measurement could be slightly biased (the radius of the calculation of the angle is important which selects a certain type of vertices). Secondly, the radius of the circle delimiting the branching area for the GIS method, is smaller from the latter one by a factor of 10. This radius was chosen in order to optimize the statistics and avoid artefacts, notably in particularly dense areas of the thallus (high quantity of matter and high number of close nodes). However, this choice could introduce a bias on the result due to the linearization intrinsic to the method which is particularly sensitive for short lengths. Lastly, another bias could be due to the fact that no discrimination between lateral and apical branching could be established  with the GIS based method, leading to take into account both cases. Moreover, there must be in the set of vertices allowing the reconstruction of the angles by the GIS method, a remainder of vertices $V_{3i}$ .
However, at 3 standard deviations, {\it i.e.} over $99.9\%$, the results remain consistent. 
 
\noindent
Overall, from both prediction and experiments we can conclude that, in our conditions, the process of angular branching allows to {\it P. anserina} to occupy as much surface as possible, and then to explore and exploit its environment in the most optimized way. All these considerations must be replaced in an {\it in vivo} context, in which the fungus is usually in competition with other organisms to occupy the colonized area and the use of available resources. So, we will consider extending the calibrated model to simulate the mycelial growth in heterogeneous environments or under various constraints, as apical branch angles respect their intrinsic property but the extension of emerged branches is influenced by external factors. Likewise, since it is possible to calibrate the simulation in time and space on the experimental data, a study of the different speeds (biological, creation of $V_1$, $V_3$, $V_{1\ell}$, $V_{3\ell}$, group velocity...) is conceivable and controllable.
A more ambitious simulation is under development. It will allow the joint analysis of local curvature in relation to branch length or lateral branch emergence.\\
To conclude, our understanding of the coordinated growth and behavior of hyphae inside the fungal network is still in its infancy. 
In this context, our work contributes to show that a reasoned reductionism on the constraints of growth, both experimentally (two-dimensional growth) and mathematically (toy-model approach) is a powerful tool to describe relevant behaviour of the mycelial growth allowing for exploring multiple hypotheses and constraints rapidly.

\section*{Acknowledgements} 
We warmly thank Sylvie Cangemi for her expert technical assistance and Frédéric Filaine for the  improvement of the experimental set-up. Clara Ledoux is supported by a PhD scholarship from the Doctoral School MTCI (ED 563).  This work was supported by the IdEx Université de Paris, ANR-18-IDEX-0001.  Special thanks to Florence for her legendary optimism.

\section*{Author contributions} 
C.L. performed biological experiments, contributed to their analysis and discussed the results. 
G.R.R. contributed to the data analysis and discussed the results. 
C.B. realized the GIS automatic approach and discussed the results.
Ch.L. contributed to the data analysis and discussed the results. 
F.C.L. and E.H. conceived the experimental setup. 
F.C.L., E.H. and P.D. conceived the ideas and the analysis, interpreted the data and wrote the article. 
P.D. performed the simulation.
All authors reviewed and amended the manuscript.

\section*{Competing interests} 
The authors declare no competing interests.

\bibliography{bibtex.bib}

\end{document}